\documentclass[twocolumn]{aastex63}
\usepackage{csvsimple}
\usepackage{amsmath}  
\usepackage{amssymb}  
\usepackage{multirow}
\usepackage{arydshln}
\usepackage{array}
\newcommand{\PreserveBackslash}[1]{\let\temp=\\#1\let\\=\temp}
\usepackage{color}

\newcommand{\kep}{{\it Kepler}}
\newcommand{\prot}{$P_\text{rot}$}
\newcommand{\protml}{$P_\text{rot, ML}$}
\newcommand{\protas}{$P_\text{rot, AutoS}$}
\newcommand{\protmcq}{$P_\text{rot, McQ14}$}
\newcommand{\sph}{$S_\text{\!ph}$}
\newcommand{\teff}{$T_\text{eff}$}
\newcommand{\logg}{$\log\, g$}
\usepackage{colortbl}
\newcommand\aastex{AAS\TeX}

\newcommand{\cpcbi}{Type~1 CP/CB}
\newcommand{\cpcbii}{Type~2 CP/CB}
\newcommand{\cpcbiii}{Type~3 CP/CB}
\newcommand{\cpcbiv}{Type~4 CP/CB}
\newcommand{\ml}{ROOSTER}
\definecolor{purple}{RGB}{102,0,204}
\definecolor{blue}{RGB}{0,10,150}

\shorttitle{\aastex\ Rotation for Kepler G and F dwarfs, and subgiants}
\shortauthors{A. R. G. Santos et al.}

\usepackage{hyperref}

\begin{document}

\title{Surface rotation and photometric activity for \textit{Kepler} targets\\
II. G and F main-sequence stars, and cool subgiant stars}

\author[0000-0001-7195-6542]{A. R. G. Santos}
\email{angela.goncalves-dos-santos@warwick.ac.uk;\\ asantos@astro.up.pt}
\affil{Department of Physics, University of Warwick, Coventry, CV4 7AL, UK}
\affil{Space Science Institute, 4750 Walnut Street, Suite 205, Boulder CO 80301, USA}

\author[0000-0003-0377-0740]{S. N. Breton}
\affil{AIM, CEA, CNRS, Universit\'e Paris-Saclay, Universit\'e de Paris, Sorbonne Paris Cit\'e, F-91191 Gif-sur-Yvette, France}

\author[0000-0002-0129-0316]{S. Mathur}
\affil{Instituto de Astrof\'isica de Canarias (IAC), E-38205 La Laguna, Tenerife, Spain}
\affil{Universidad de La Laguna (ULL), Departamento de Astrof\'isica, E-38206 La Laguna, Tenerife, Spain}

\author[0000-0002-8854-3776]{R. A. Garc\'{i}a}
\affil{AIM, CEA, CNRS, Universit\'e Paris-Saclay, Universit\'e de Paris, Sorbonne Paris Cit\'e, F-91191 Gif-sur-Yvette, France}


\begin{abstract}
Dark magnetic spots crossing the stellar disc lead to quasi-periodic brightness variations, which allow us to constrain stellar surface rotation and photometric activity. The current work is the second of this series \citep[][Paper~I]{Santos2019a}, where we analyze the \kep\ long-cadence data of 132,921 main-sequence F and G stars and late subgiant stars. Rotation-period candidates are obtained by combining wavelet analysis with autocorrelation function. Reliable rotation periods are then selected via a machine learning (ML) algorithm \citep{Breton2021}, automatic selection, and complementary visual inspection. The ML training data set comprises 26,521 main-sequence K and M stars from Paper~I. To supplement the training, we analyze in the same way as Paper~I, i.e. automatic selection and visual inspection, 34,100 additional stars. We finally provide rotation periods \prot\ and associated photometric activity proxy \sph\ for 39,592 targets. Hotter stars are generally faster rotators than cooler stars. For main-sequence G stars, \sph\ spans a wider range of values with increasing effective temperature, while F stars tend to have smaller \sph\ values in comparison with cooler stars. Overall for G stars, fast rotators are photometrically more active than slow rotators, with \sph\ saturating at short periods. The combined outcome of the two papers accounts for average \prot\ and \sph\ values for 55,232 main-sequence and subgiant FGKM stars (out of 159,442 targets), with 24,182 new \prot\ detections in comparison with \citet{McQuillan2014}. The upper edge of the \prot\ distribution is located at longer \prot\ than found previously.
\end{abstract}

\keywords{stars: low-mass -- stars: rotation -- stars: activity -- starspots -- techniques: photometric -- methods: data analysis -- catalogs}
%

\section{Introduction}

The measurement of the rotation of solar-like stars (i.e. stars with an external convective envelope) has been at the center of many studies in stellar physics. Internal rotation modifies the mixing of elements inside stars. During the main sequence, rotation refuels the Hydrogen content in the stellar core from the upper layers, extending the main-sequence lifetime of faster rotators compared to slower rotators \citep[e.g.][]{Aerts2019}. Asteroseismology, i.e. the study of stellar oscillations \citep[e.g.][]{Garcia2019}, provides a unique way to infer stellar properties, including internal rotation. Unfortunately, only for evolved stars it has been possible to measure the core rotation \citep[e.g][]{Beck2012,Deheuvels2012,Deheuvels2014,Gehan2018,Mosser2018}. For main-sequence solar-like stars, asteroseismology can only provide reliable constraints on rotation of the outermost layers and for a small number of stars \citep[e.g.][]{Benomar2015,Benomar2018}. Alternatively, surface rotation can be measured from long-term brightness variations due to magnetic features co-rotating with the stellar surface.

As shown by \citet{Skumanich1972}, there is a tight relation between stellar age and surface rotation for low-mass solar-like stars: stars spin-down as they evolve due to magnetic braking. This inspired the so-called gyrochronology \citep[e.g.][]{Barnes2003,Barnes2007,Meibom2011a,Meibom2011b,Meibom2015,Mamajek2008}, which could allow us to estimate stellar ages for large samples of field stars with high precision. However, the validity of the Skumanich spin-down law throughout the main sequence has been subject to debate. On the one hand, the recent results by \citet{Lorenzo-Oliveira2019,LorenzoOliveira2020} are consistent with a steady spin-down, supporting gyrochronology as reliable. On the other hand, other studies invoke a weakening of the magnetic braking to explain the discrepancy found between the asteroseismic ages and those predicted by gyrochronology \citep[e.g.][]{Angus2015,vanSaders2016,Metcalfe2019,Hall2021}.
Therefore, the determination of reliable rotation periods is crucial to understand the spin-down evolution and derive precise stellar ages where applicable.

Thanks to the advent of planet-hunting space missions, like CoRoT \citep[Convection, Rotation et Transits plan\'etaires;][]{Baglin2006a}, \kep\ \citep{Borucki2010}, K2 \citep{Howell2014}, and TESS \citep[Transiting Exoplanets Survey Satellite;][]{Ricker2014}, such measurements are possible for an extraordinary number of targets \citep[e.g.][]{Mosser2009,Nielsen2013,Garcia2014,McQuillan2014,Santos2019a,Reinhold2020,Gordon2021}.

In this work, we focus on the analysis of the four-year \kep\ data, which correspond to the longest continuous high-precision photometric survey obtained so far for hundreds of thousands of stars. In the near future, no other on-going or planned space mission will provide a better dataset in terms of continuous long-term photometric monitoring. Here, we estimate rotation periods following the same methodology as in \citet[][hereafter Paper~I]{Santos2019a}, which combines three different rotation diagnostics for different calibrated time series. Paper~I reported the detection of new rotation periods for 4,431 stars in comparison to \citet[][hereafter McQ14]{McQuillan2014} for main-sequence K and M stars, according to the stellar properties of \citet{Mathur2017}. The current work, the second of this series, extends the analysis to solar-like stars of spectral type G and F as well as solar-like subgiants.  Because the target sample of this work is several times larger than that of Paper~I, we use a machine learning algorithm \citep[\ml\ - Random fOrest Over STEllar Rotation;][]{Breton2021} to reduce the amount of visual inspections with respect to those carried out in Paper~I.

The manuscript is organized as follows. Section~\ref{sec:datasample} describes the target selection and data calibration. Although the original target selection was done according to the stellar parameters from \citet{Mathur2017}, we adopt for the remainder of the analysis the recent stellar properties catalog by \citet{Berger2020}. For continuity of Paper~I, the title of the current work reflects the stellar classification by \citet{Mathur2017}, i.e. a target sample of F and G main-sequence stars and subgiants. Nevertheless, the majority of the targets in the current analysis is indeed consistent with mid-F to G main-sequence or subgiant stars also according to \citet{Berger2020}. The rotation pipeline, used to retrieve rotation-period candidates, and the photometric magnetic activity proxy are described in Sects.~\ref{sec:rotation} and \ref{sec:sph}. Reliable rotation periods are then selected through the implementation of a machine learning algorithm, automatic selection, and supplementary visual inspection (Sect.~\ref{sec:protsel}). The results are finally presented and discussed in Sects.~\ref{sec:results} and \ref{sec:conclusions}.

\section{Data preparation and sample selection}\label{sec:datasample}

\subsection{Data preparation}\label{sec:data}

In  this work, we analyze long-cadence ($\Delta t=29.42$ min) data obtained by the \kep\ main mission. We use KEPSEISMIC light curves\footnote{KEPSEISMIC time-series are available at MAST via \dataset[https://doi.org/10.17909/t9-mrpw-gc07]{https://doi.org/10.17909/t9-mrpw-gc07}.} \citep{Garcia2011}, which are optimized for seismic studies but are also appropriate for rotational analysis (e.g. Paper~I). The light curves are obtained from \kep\ pixel-data with custom apertures, which are typically larger than those used for PDC-MAP \citep[Presearch Data Conditioning - Maximum A Posteriori; e.g][]{Jenkins2010,Smith2012,Stumpe2012} data products. See Paper~I for further details on the KEPSEISMIC apertures. The resulting light curves are then processed by KADACS \citep[\kep\ Asteroseimic data analysis and calibration Software;][]{Garcia2011}. In addition to correcting for outliers, jumps, drifts, and discontinuities at the \kep\ quarter edges, KADACS implements in-painting techniques \citep{Garcia2014a,Pires2015} to fill gaps shorter than 20 days using a multi-scale discrete cosine transform. Finally, the light curves are high-pass filtered at 20, 55, and 80 days. This way, for each star we have three KEPSEISMIC light curves. While the filters with short cutoff period deal better with \kep\ instrumental effects than the filters with long cutoff period, they can also filter the intrinsic stellar rotational modulation. Therefore, we opt to use and compare the results for the three different filters. Note that it is possible to retrieve rotation periods longer than the cutoff period, because the transfer function of a given filter slowly approaches zero at twice the cutoff period. Furthermore, we also analyze PDC-MAP light curves for Data Release 25 in order to ensure that the retrieved period from KEPSEISMIC data is not a consequence of the large custom apertures, being the result of photometric pollution by a nearby star. Aside from the difference in the KEPSEISMIC and PDC-MAP apertures, PDC-MAP light curves are often filtered at 20 days, which leads to a clear bias on the rotation results (Paper~I).


\subsection{Sample selection}\label{sec:sample}

The target samples for Paper~I and for the current study were originally defined with the \kep\ Stellar Properties Catalog for Data Release 25 \citep[][hereafter DR25]{Mathur2017}, which was the latest update to the \kep\ stellar properties available at the time for Paper~I. In Paper~I, we analyzed \kep\ stars that were classified as K and M main-sequence stars according to DR25 (cooler than $T_\text{eff}=5200$ K). Here, we analyze the remainder of the \kep\ main-sequence and subgiant targets expected to be solar-like stars, i.e. stars with convective outermost layers. According to DR25, the current work focuses on main-sequence stars from spectral type mid-F to G, as well as subgiant stars from spectral type mid-F to K. Nonetheless, for the current analysis we decide to embrace the new update to the \kep\ stellar properties, i.e. the recent {\it Gaia}-\kep\ Stellar Properties Catalog \citep[][hereafter B20]{Berger2020}. As follows, in the light of B20, the current classification differs from that considered in Paper~I (see below). The detailed comparison between the two stellar properties catalogs is presented in B20. In Appendixes~\ref{app:Fstars}-\ref{app:allB20}, we discuss some of the differences relevant for the targets in our sample.

\begin{figure*}
    \centering
    \includegraphics[width=\hsize]{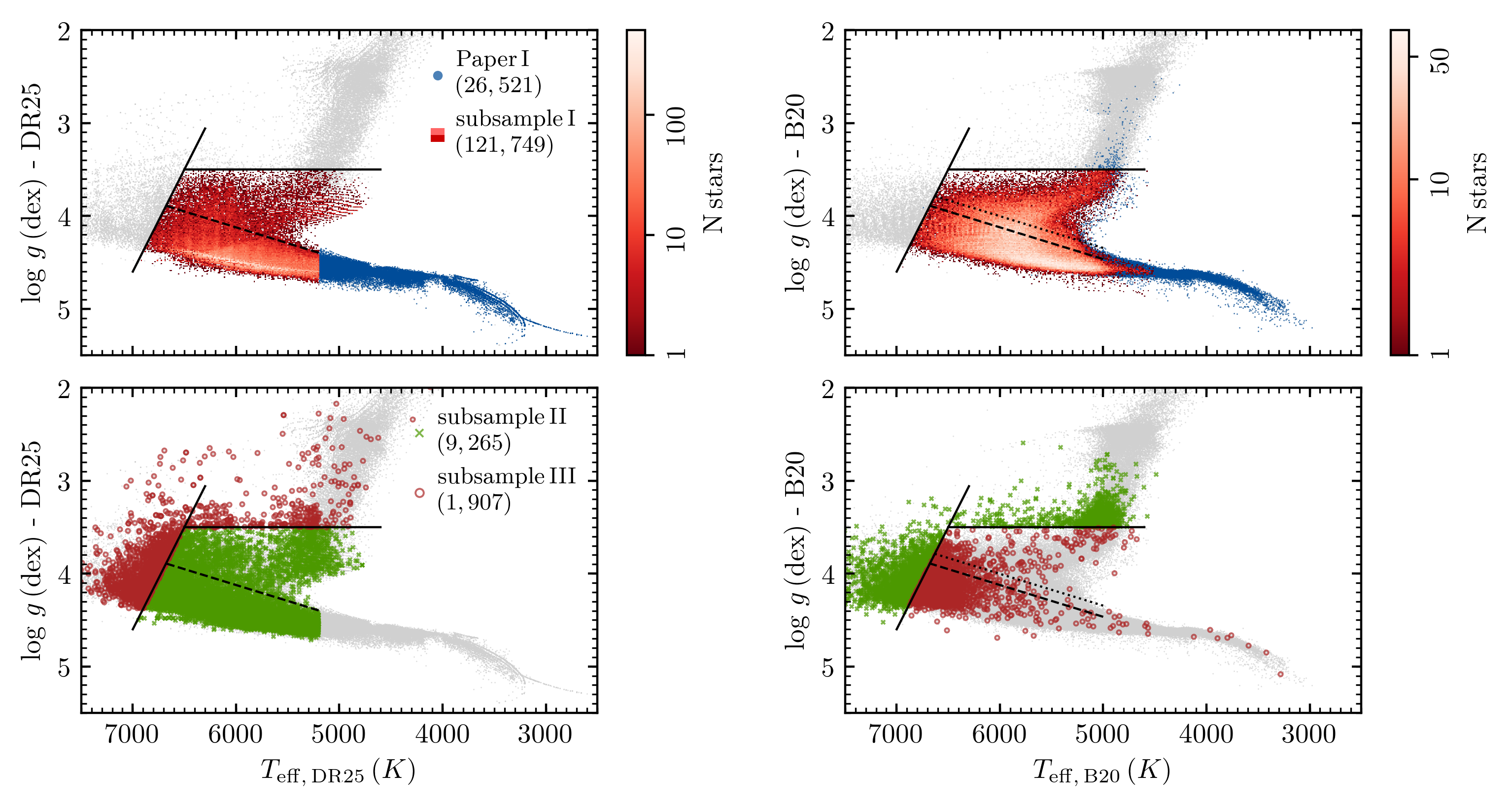}
    \caption{Surface gravity-effective temperature diagram for the targets considered in the rotational analysis (colored data points) according to DR25 (left) and B20 (right). The target sample of the current study consists of subsamples~I (top; shades of red), II (bottom; green crosses), and III (bottom; red circles). Subsample~I, color coded by the number of stars in each bin, corresponds to targets that are solar-like main-sequence or subgiant solar-like stars in both catalogs (DR25 and B20). Subsample~II targets are solar-like stars in DR25 but not in B20. Subsample~III targets are solar-like stars in B20 but not in DR25. The blue dots mark the target sample of Paper~I, which is used as part of the training set in the current analysis. The solid lines mark the instability strip and surface gravity cut adopted to select main-sequence and subgiant stars, while the dashed line marks the separation between main-sequence and subgiant phases. For reference all \kep\ targets are plotted in gray.}
    \label{fig:HR}
\end{figure*}

Figure~\ref{fig:HR} shows the target samples of Paper~I and the current study according to DR25 (left) and B20 (right).
We adopt the classical instability strip \citep[diagonal solid line in Fig.~\ref{fig:HR};][]{Bowman2018,Dupret2005} to select the targets expected to have convective envelopes. To avoid potential red giants we consider a flat cut at $\log\,g=3.5$ (horizontal solid line). We then remove contaminants from the target sample: $\delta$ Scuti, $\gamma$ Doradus, and hybrids \citep{Uytterhoeven2011,Bradley2015,vanReeth2018,Murphy2019,Li2019,Li2019b}; RR Lyrae stars \citep[][Szab\'o et al. in prep.]{Benko2010,Nemec2011,Nemec2013}; misclassified red giant stars (Garc\'ia et al. in prep. and references therein); and eclipsing binaries \citep[Villanova \kep\ Eclipsing Binary Catalog;][]{Kirk2016,Adbul-Masih2016}. In total we remove 8,209 known contaminants that are within the parameter space of the target sample of this work.

The top panels of Fig.~\ref{fig:HR} show in color the targets that are main-sequence or subgiant stars in both stellar properties catalogs (DR25 and B20). The target sample of Paper~I, used in this work as part of the training for the machine learning algorithm \citep{Breton2021}, is plotted in blue. The targets colored in shades of red belong to the target sample for the rotational analysis of this work (121,749 targets -- subsample~I), after removing the contaminants listed above. The bottom panels of Fig.~\ref{fig:HR} show the targets whose classification, in terms of being solar-like, disagrees between the two catalogs that are still considered in the rotational analysis: main-sequence or subgiant solar-like stars in DR25 but not in B20 (green; 9,265 targets -- subsample~II); main-sequence or subgiant solar-like stars in B20 but not in DR25 (red; 1,907 targets -- subsample~III). In total, the target sample considered for the rotational analysis in this work comprises 132,921 stars (subsamples~I, II, and III).  

For the remainder of the analysis, we prioritize the stellar properties from B20, which are listed in Tables~{\color{blue}1} and {\color{blue}2}. When not available (part of subsample~III), we adopt the stellar properties from DR25. Accordingly, in Tables~{\color{blue}1} and {\color{blue}2}, we also provide a flag indicating the stellar properties source.

In Sect.~\ref{sec:results} we present the results for the targets in subsample I according to their spectral type and evolutionary stage. There, the targets are split following B20. To separate main-sequence from subgiant stars, we take the transition between main-sequence and the subgiant branch from evolutionary tracks for solar metallicity and different stellar masses obtained with MESA \citep[Modules for Experiments in Stellar Astrophysics;][and references therein]{Paxton2018} and fit a linear relation shown by the dashed line in Fig.~\ref{fig:HR}. This cut leads to 21,665 subgiant stars in subsample~I according to the originally adopted DR25. However, this cut is not appropriate for B20 as the main-sequence is slightly shifted towards small \logg\ values. Therefore, we shift the line by -0.12 dex in \logg\ (dotted line in the right-hand panels). Using this cut and B20 parameters leads to similar statistics, now with 22,843 subgiant stars. We consider the boundary between main-sequence G and F stars at $T_\text{eff}=6000$ K.

In addition to known contaminants reported in the literature and described above (listed in Table~{\color{blue}2}), there are a number of other contaminants that may still remain in the data. Here, we do not provide rotation period for light curves with photometric pollution (e.g. when the signal is only present every four \kep\ quarters) or multiple signals. These targets are listed in Table~{\color{blue}2} with the respective flag. Multiple signals can result from photometric pollution by background stars or from unresolved multiple systems. Determining the source of the multiple signals is beyond the scope of this work. Thus, we do not consider these targets in the subsequent analysis. 

Following the approach in Paper~I we flag CP/CB (Classical Pulsator/Close-in Binary) candidates. \cpcbi\ candidates show high-amplitude brightness variations, stable and fast beating patterns, and/or a large number of harmonics. In Paper~I, we discuss the possibility of these targets being tidally synchronized binaries, which are common among rapidly rotating \kep\ targets \citep{Simonian2019,Angus2020}. We provide rotation periods (Table~{\color{blue}1} with the proper flag) for these targets as the signal can still be related to rotation but not of single stars. The signal of \cpcbii\ candidates resembles that of contact binaries \citep[e.g.][]{Lee2016,Colman2017}. \cpcbiii\ candidates are $\delta$ Scuti and/or $\gamma$ Doradus candidates or alternatively polluted by a nearby star of this type. Additionally, in this work we flag another potential type of CP/CB candidates. The signal of \cpcbiv\ candidates resembles that of heartbeat stars or close binaries with tidally excited oscillations \citep[e.g.][]{Guo2020}. The signatures of Type~2-4 CP/CB candidates can be mistakenly selected as rotation and are identified during visual examination. We do not provide periods for Type~2-4 CP/CB candidates, instead these are listed in Table~{\color{blue}2} with the respective flag. 

\section{Surface rotation and photometric magnetic activity}

\subsection{Rotation-period candidates}\label{sec:rotation}

In this section, the methodology used to estimate the rotation-period candidates from the stellar brightness variations is briefly described. For more details see Paper~I and \citet{Ceillier2016,Ceillier2017}.

Our rotation pipeline combines a time-frequency analysis and the autocorrelation function (ACF). Using artificial data, \citet{Aigrain2015} concluded that such combination of different rotation diagnostics, together with a performant time series preparation, provides the most complete set of reliable rotation-period estimates \citep[see also Appendix B in ][]{Breton2021}. Compared with McQ14, in addition to a different rotation analysis, we use the full length of the \kep\ observations and we obtain and calibrate our own light curves (KEPSEISMIC) using different high-pass filters (see Sect.~\ref{sec:data}). As a result, for M and K main-sequence stars (Paper~I), we were able to recover rotation periods for 4,431 targets for which McQ14, using ACF alone, did not report a rotation period.

\begin{figure}[h!]
\includegraphics[width=\hsize]{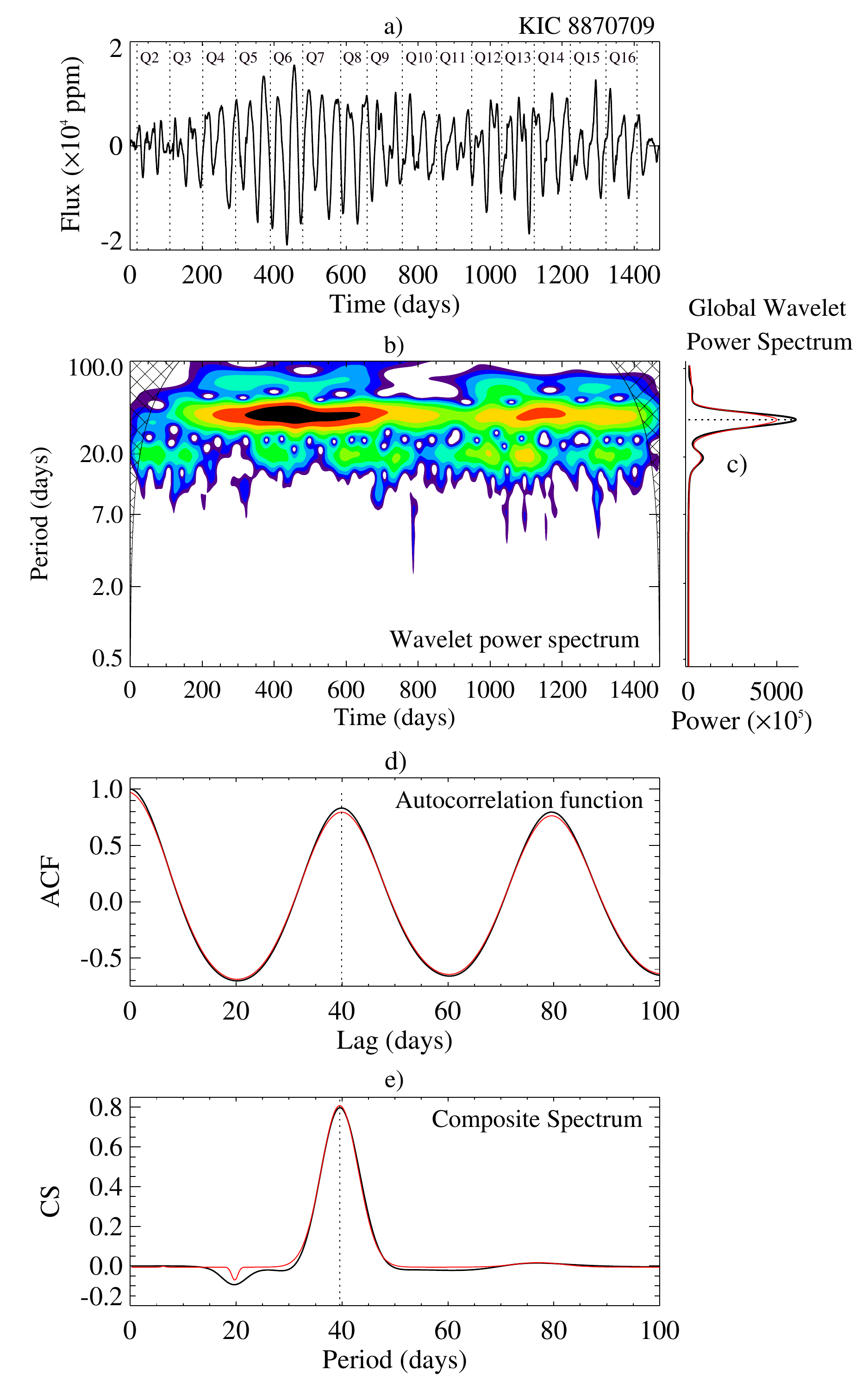}
\caption{Example of the rotation pipeline output for KIC~8870709. a) 55-day filtered KEPSEISMIC light curve. b) Wavelet power spectrum: black corresponds to high power and blue to low power. The black crossed area represents the cone of influence. c) GWPS (black) and corresponding best fit with multiple Gaussian functions (red). d) ACF (black) of the light curve and smoothed ACF (red). e) CS (black) and respective fit with multiple Gaussian functions (red). The black dotted lines mark the three rotation-period estimates (from GWPS, ACF, and CS).}\label{fig:a2ztool}
\end{figure}

Our rotation analysis retrieves three rotation-period estimates for each light curve, i.e. nine estimates per star. We obtain the first estimate from the global wavelet power spectrum (GWPS; panels b) and c) in Fig.~\ref{fig:a2ztool}), which results from the wavelet decomposition \citep{Torrence1998}. The wavelet decomposition was first adapted for the analysis of stellar light curves by \citet{Mathur2010}, who adopted the correction by \citet{Liu2007}. Following the method by \citet[][see also \citet{Garcia2014}]{McQuillan2013a}, we obtain the second period estimate from the autocorrelation function of the light curve (ACF; panel d) in Fig.~\ref{fig:a2ztool}). Finally, the third period estimate is provided by the composite spectrum (CS; panel c) in Fig.~\ref{fig:a2ztool}) which is the product between the normalized GWPS and the normalized ACF \citep[for its first application see][]{Ceillier2016,Ceillier2017}. As the common periods between GWPS and ACF are highlighted by the CS, this diagnostic allows us to better distinguish the stellar rotation signals from false positives, such as instrumental modulations. 

For the final period estimate we prioritize the value provided by the GWPS, whose uncertainty is typically large accounting for the uncertainty on the period determination and partially for differential rotation. 

\subsection{Photometric Magnetic Activity Proxy}\label{sec:sph}

Once we have the rotation-period candidates for the targets, we can obtain the photometric activity proxy.

The photometric magnetic activity proxy \sph \citep{Garcia2010,Mathur2014} is a measure of the amplitude of the spot modulation in the light curves. \sph\ is computed as the standard deviation of light curve segments of length 5 times the rotation-period candidate. We correct \sph\ for the photon noise following the approach by \citet{Jenkins2010}.

The photon noise correction can lead to negative \sph\ values when the rotational modulation is absent or its amplitude is small. Note that, if rotational modulation is not detected, the \sph\ value does not provide a proxy for magnetic activity. Nevertheless, we feed the machine learning algorithm (Sect.~\ref{sec:ML}) with these \sph\ values: one per rotation-period candidate.

After selecting the targets with rotational modulation by following the steps described below, for those with an over-corrected \sph\ we apply a different individual correction computed from the high-frequency noise in the power density spectrum. These targets account for less than 0.3\% of those with final \prot\ estimate (Table~{\color{blue}1}).

We also note that the \sph\ can be a lower limit of the true photometric activity level, depending on, for example, stellar inclination angle and spot latitudinal distribution. Nevertheless, \sph\ has been shown to be a good proxy for solar and stellar magnetic activity \citep[][]{Salabert2016a,Salabert2017}.

\pagebreak

\subsection{Rotation-period selection}\label{sec:protsel}

In Paper~I, the selection of reliable rotation periods was made essentially in two steps. Periods were automatically selected if the rotation-period estimates agree between different diagnostics and filters, and if the height of the respective rotation peaks is larger than a given threshold (for details see Paper~I). In the second step, for the targets whose period was not automatically selected, we proceeded with visual examination of the light curves, results from the rotation pipeline, and power spectrum density. We visually inspected about 60\% of the target sample of Paper~I comprised of 26,521 main-sequence K and M stars, according to DR25 (note that for the current analysis we adopt B20). Here, we analyze the remainder of the targets observed by \kep\ expected to be main-sequence or subgiant solar-like stars. The target sample of this work is then comprised of 132,921 targets. Therefore, it is crucial to reduce the number of required visual checks.

In order to do so, here we use a machine learning algorithm, ROOSTER, to identify targets with rotational modulation and select the respective period. ROOSTER and its validation are described in detail in \citet[][]{Breton2021} and summarized in Sect.~\ref{sec:ML}. In the context of this work, the main goal of the implementation of machine learning is to efficiently select reliable rotation periods while reducing the required amount of visual inspection. To that end, we also need to supply the machine learning algorithm with a proper training set (Sect.~\ref{sec:training}). 

\subsubsection{Training set}\label{sec:training}

\begin{figure}
    \centering
    \includegraphics[width=\hsize]{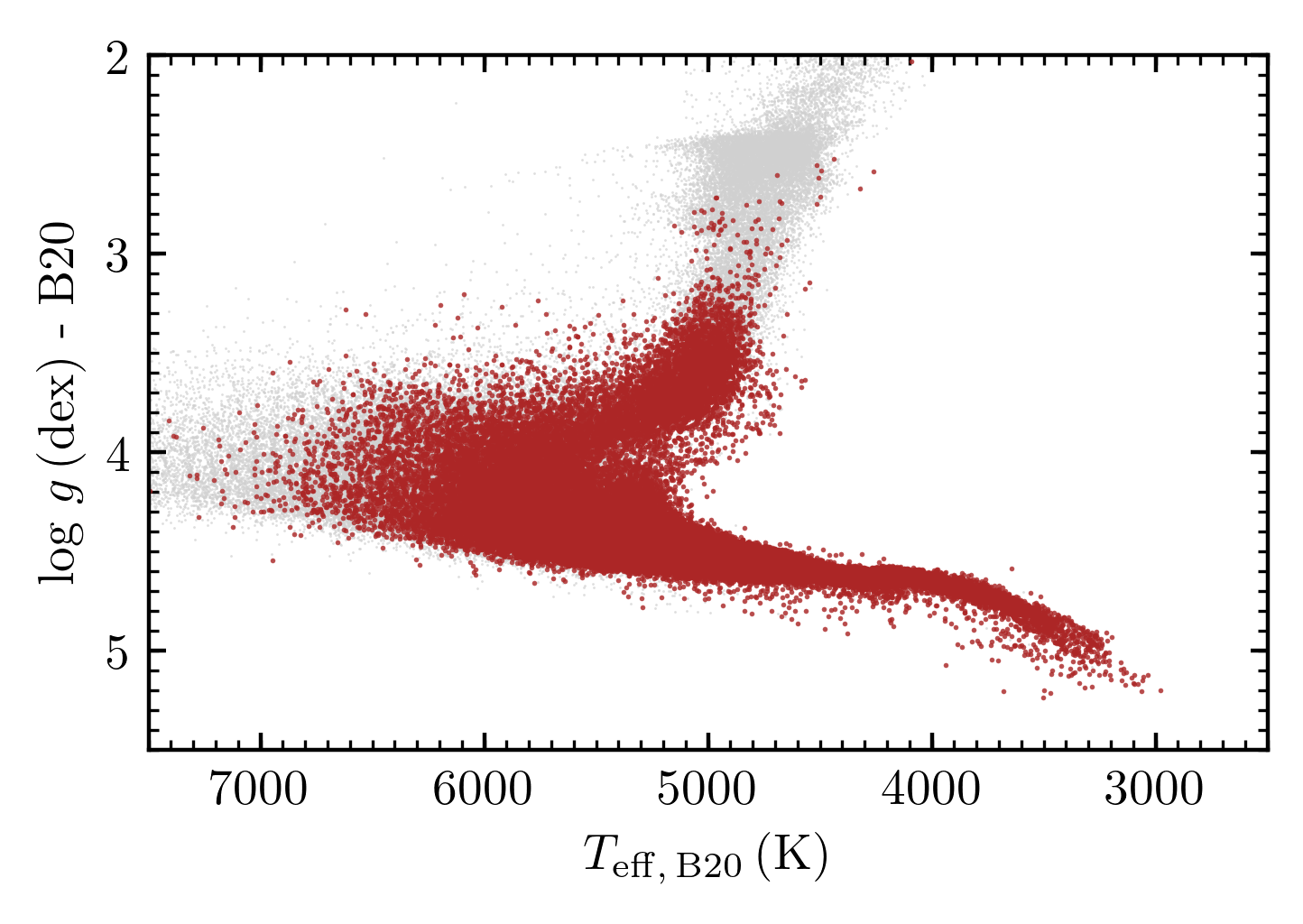}
    \caption{Surface gravity-effective temperature diagram for the targets in the training set (red). The training set comprises the targets of Paper~I and 34,100 additional targets from subsamples I, II, and III. The preparation of the target sample used automatic selection and visual inspection and validation (see main text).} For reference, all \kep\ targets are plotted in gray.
    \label{fig:hr_training}
\end{figure}

For the training set of ROOSTER, we use the 26,521 solar-like stars from Paper~I. At the time, these targets were classified as K and M main-sequence stars (left top panel of Fig.~\ref{fig:HR}). According to B20 (right-hand panel of Fig.~\ref{fig:HR}), the latest stellar properties catalog, most of the stars in Paper~I (95.4\%) are indeed cool solar-like stars with $T_\text{eff}<5200$ K. To complement the training set, namely to account for the full range of target effective temperatures, we analyze 34,100 stars from subsamples~I-III (Fig.~\ref{fig:hr_training}) in the same manner as the targets of Paper~I, i.e. through automatic selection and visual inspection. The rotational signal of hotter stars, in particular F stars, differs from that of cooler stars. Thus, to avoid bias and properly train the machine learning tool, it is important to consider a diverse training set. 

\textit{Automatic selection} - Rotation periods are automatically selected if there is agreement between the \prot\ estimates from the different diagnostics (GWPS, ACF, CS) and KEPSEISMIC light curves obtained with three different filters. Additionally, we impose a height threshold for the ACF and CS rotation peaks. See Paper~I for details.

\textit{Visual inspection} - The light curves (three KEPSEISMIC and one PDC-MAP; see Sect.~\ref{sec:data}), power density spectra, and rotation diagnostics of all fast rotators ($P_\text{rot}<10$ days), slow rotators ($P_\text{rot}>60$ days), and targets for which the rotation period is not automatically selected are visually inspected. 

The final training set is composed of 60,621 targets (Fig.~\ref{fig:hr_training}): 29,563 targets with rotation-period estimate, including \cpcbi\ candidates; and 31,058 targets without rotation-period estimate. This leaves 98,821 targets to be analyzed by the machine learning algorithm. Note that the targets analysed in Paper~I (26,521) are not part of the target sample of the current work, being only part of the training.


\subsubsection{Machine learning algorithm: \ml}\label{sec:ML}

\citet{Breton2021} developed a machine learning tool, \ml, to select reliable rotation periods from the output of the rotation pipeline (Sect.~\ref{sec:rotation}). For each target, ROOSTER's input parameters are the nine rotation-period candidates (Sect.~\ref{sec:rotation}) and respective \sph\ values (Sect.~\ref{sec:sph}), additional control parameters from the rotation pipeline (e.g. ACF and CS peak heights), stellar fundamental properties, FliPer metric \citep{Bugnet2018}, and observation parameters (e.g. \textit{Kepler} magnitude, observation length).
\ml\ employs three random forest classifiers, each one dedicated to a specific task. The first classifier selects stars with rotational modulation from the target sample. For those selected stars, the second classifier provides a flag, which identifies \cpcbi\ candidates (see Sect.~\ref{sec:sample}). Finally, for the same selected stars, the third classifier chooses the rotation period from the nine provided estimates (Sect.~\ref{sec:rotation}). 

The validation of \ml\ is presented in detail by \citet{Breton2021}, where the target sample was that of Paper~I, which comprises mostly KM stars (see Appendix~\ref{app:MLtrainingset} for the full training set, i.e. stars of spectral type from mid-F to M). For the cool solar-like targets, the initial \ml's global accuracy was 92.1\%. In spite of the good yield, the results can still be improved through supplementary visual inspection. The main source of the \ml's confusion are the targets with high-amplitude second harmonics of the rotation period, for which several of the nine period estimates provided as input parameters are half of the true rotation period. Another group of intricate targets corresponds to light curves with long-term instrumental modulations. Often the period selected by \ml\ for these targets is between 38 and 60 days. In Sect.~\ref{sec:vischecks}, we then carry out a number of steps to identify the potential \ml\ period misselections. In fact for the cool solar-like stars, \citet{Breton2021} concluded that the methodology's accuracy can be improved from 92.1\% to 96.9\% by identifying relevant targets for visual inspection.

Note that for the current analysis, which focuses on hotter solar-like stars than those in Paper I and \citet{Breton2021}, we have doubled the training set (Sect.~\ref{sec:training} and Appendix~\ref{app:MLtrainingset}). We currently train \ml\ with stars of spectral type mid-F to M.

\subsubsection{Supplementary visual inspection}\label{sec:vischecks}

As mentioned above, during the development and validation of the machine learning tool, we have identified problematic groups of targets for which \ml\ has difficulty on selecting the correct rotation period. For this reason the machine learning \prot\ selection is complemented with additional checks and visual inspection. Below we describe the most relevant groups of targets, rather than describe the full assessment from the visual inspection.

{\it Targets with missing input parameters for the machine learning tool:} For a fraction of stars ($\sim 1\%$ of the sample considered for the machine learning), \ml\ is unable to provide an assessment, because some of the input parameters, namely those related to the ACF, are not determined. We visually check these targets. A significant part of the targets ($\sim56\%$) do not show rotational modulation and only for $\sim10\%$ we provide a final rotation period (Table~{\color{blue}1}).

{\it Comparison with the automatic selection:} We first compare the \prot\ values for targets common to the ML selection (\protml) and automatic selection (\protas). \protas\ always corresponds to the period recovered from the wavelet analysis. We visually inspect all the targets in disagreement, as well as the targets with \protas\ but no \protml. The targets for which the machine learning and the automatic selection disagree usually correspond to targets with high-amplitude second harmonics for which \ml\ selects half of the rotation period ($\sim 93\%$ of the targets in disagreement). Table~\ref{tab3} summarizes the \prot\ values finally selected after the visual inspection.

{\it Comparison with the literature:} Next, we cross-check the targets with \protas\ and/or \protml\ with the \prot\ values reported by McQ14 (\protmcq). Similarly to the previous step, we visually inspect all the targets with \protas\ and/or \protml\ in disagreement with \protmcq. Also, we visually check the targets with \protmcq\ that were not automatically selected or selected by the machine learning, with exception of the known contaminants (Sect.~\ref{sec:sample}). For the targets with both \protas\ and \protml\ in disagreement with \protmcq, our \prot\ estimates are correct for $\sim 47\%$ of the targets. For the remainder of the targets, their light curves often show photometric pollution ($\sim29\%$) or are CP/CB candidates ($\sim 18\%$). The rotation periods selected solely by the machine learning (i.e. without \protas) in disagreement with McQ14 usually correspond to half of the true rotation periods ($\sim84\%$ of the targets with discrepant \protml\ and \protmcq). Photometrically polluted light curves contribute to $\sim 13\%$ of the disagreement between \protml\ and \protmcq. Targets with \prot\ reported in McQ14 but not in this work correspond mostly to CP/CB candidates, light curves with instrumental modulation or photometrically polluted, and known contaminants (e.g. red giants, $\delta$~Scuti, $\gamma$~Doradus). These targets are listed in Table~{\color{blue}2}.

We then proceed to identify additional wrongly selected \protml\ or rotation periods of targets that may have been missed by \ml. The visual inspections described in the subsequent paragraphs concern targets for which \prot\ was not automatically selected or reported by McQ14. 

{\it Mistaken filter choice:} We verify whether the proper filter is being selected, i.e. 20-day filter for $P_\text{rot}<23$ days, 55-day filter for $23\leq P_\text{rot}<60$ days, and 80-day filter for $P_\text{rot}>60$ days (see Paper~I for details). The objective of this choice is to ensure that the impact from instrumental modulations on the \sph\ value is minimized, while \prot\ is unaffected by the filtering. If the proper filter was not chosen by the machine learning, but the GWPS rotation period agrees within $15\%$ of \protml, we automatically change the \prot\ value to that retrieved by the GWPS in the proper filter (Table~{\color{blue}1}). If the \protml\ and the \prot\ value in the proper filter disagree, we proceed for visual examination to decide whether there is a rotational signal and decide on the correct period. This disagreement often results from the presence of long-term instrumental modulations. Thus, \ml\ is giving preference to the \prot\ results from the 20-day filtered light curves. If the filtering does not affect \prot, \protml\ is kept. We corrected the \prot\ values for $\sim 24\%$ of the targets in this conditions, while $\sim 17\%$ of the targets were demoted to no \prot\ detection (Table~{\color{blue}2}).

{\it Potential CP/CB candidates:} \ml\ also flags \cpcbi\ candidates, which are typically fast rotators with $P_\text{rot}<7$ days (see Paper~I). Thus, we visually check the \cpcbi\ candidates with $P_\text{rot}>7$ days: the \cpcbi\ flag is removed for about 29\% of the targets. Secondly, to ensure that we are not providing rotation periods for \cpcbii\ candidates, we visually inspect the targets with $P_\text{rot}\leq 1.6$ days (for targets in Paper~I, the periodicity of the signal of all \cpcbii\ candidates is shorter than 1.6 days). $\sim 17\%$ of these targets are actually affected by photometric pollution.

{\it Potential instrumental modulation:} As mentioned in Sect.~\ref{sec:ML}, the results from \ml\ are affected by some confusion with instrumental-related modulations. Therefore, we visually inspect the targets with ML \prot\ longer than 38 days. About $\sim 87\%$ of these targets belong indeed to the rotation table but for $\sim 11\%$ of them we choose a different \prot\ after the visual inspection. 

{\it Potential harmonics:} Another problematic group of targets for \ml, identified in \citet[][Sect.~\ref{sec:ML}]{Breton2021}, corresponds to targets with high-amplitude second harmonics. For this type of targets, half of the rotation period may be reported \citep[see for example discussion in ][]{McQuillan2013a,McQuillan2014}. Therefore, we visually check the targets for which one or more \prot\ estimates (9 for each target) are the double of \protml. The machine learning algorithm had wrongly selected the harmonic for $\sim28\%$ of these targets. In particular, for the targets with at least three \prot\ estimates being the double of \protml, \ml\ had selected half of the correct \prot\ for about 58\% of the targets. The targets mentioned here exclude the half \prot\ already identified in previous steps. 

{\it \prot\ probability between 0.4 and 0.8:} In \citet{Breton2021}, we found that there is an area of confusion where a significant number of targets without \protml\ exhibit rotational signal, and vice-versa. This corresponds to targets with a \protml\ probability between 0.4 and 0.8, which we visually check. We determined that $\sim 61\%$ of the targets without \protml\ have rotational modulation and selected the respective period. $\sim20\%$ of the targets in this visual inspection with \protml\ are corrected or demoted to no \prot\ detection. 

{\it Short light curves:} Finally, we visually check light curves shorter than five \kep\ Quarters with \protml\ estimate, to ensure that the ML algorithm decision is correct. From the targets left to visual check in this step, \protml\ is correct for $\sim 96\%$. 

In total, we visually checked $\sim26\%$ of the 98,821 targets (i.e. 25,477 targets) analyzed by the machine learning algorithm. This corresponds to a significant decrease in the visual inspections in comparison with Paper~I (e.g. Sect.~\ref{sec:protsel}).


\begin{figure*}\centering
\includegraphics[height=11cm,angle=90]{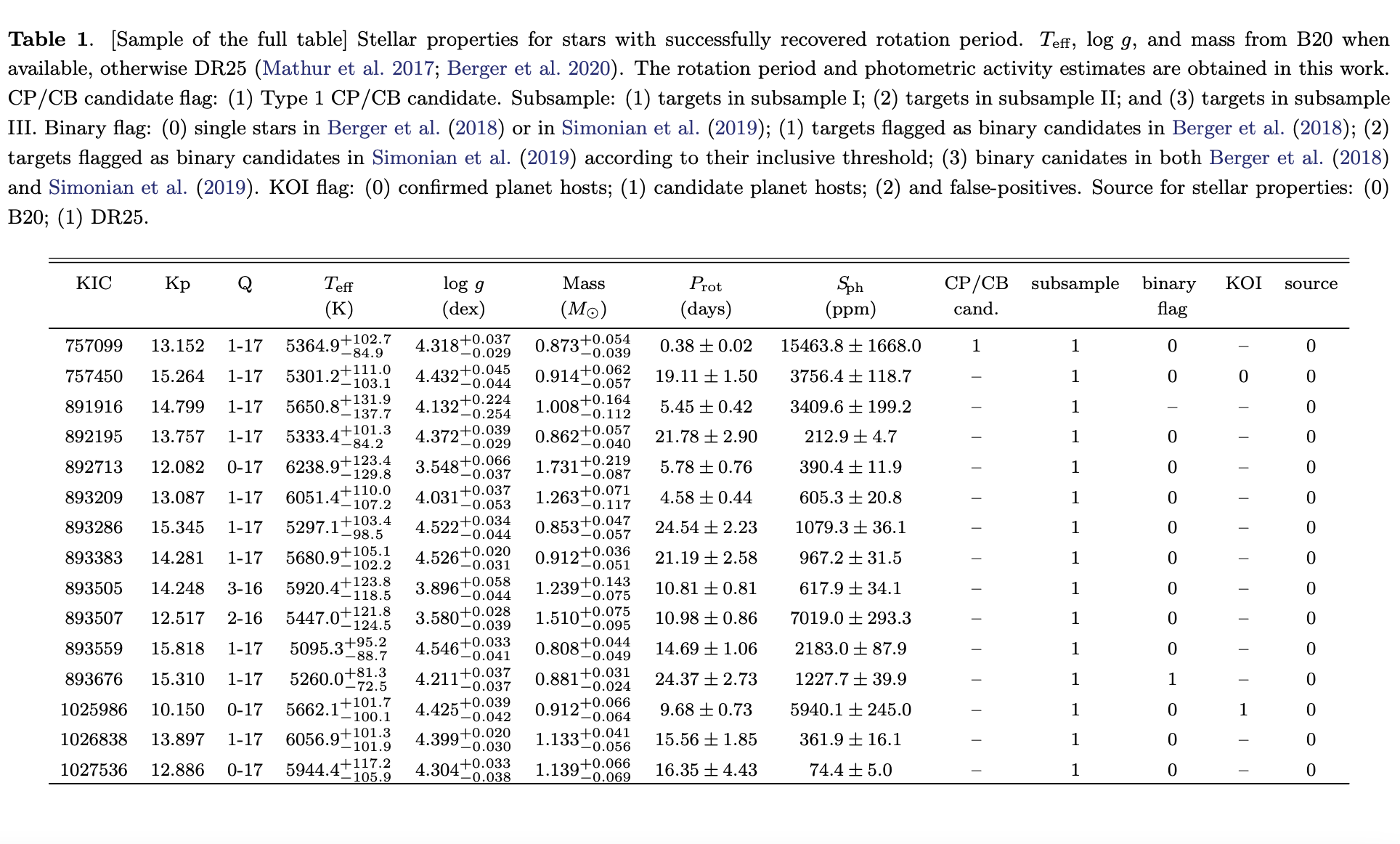}\vspace{0.6cm}
\end{figure*}

\begin{figure*}\centering
\includegraphics[height=11.5cm,angle=90]{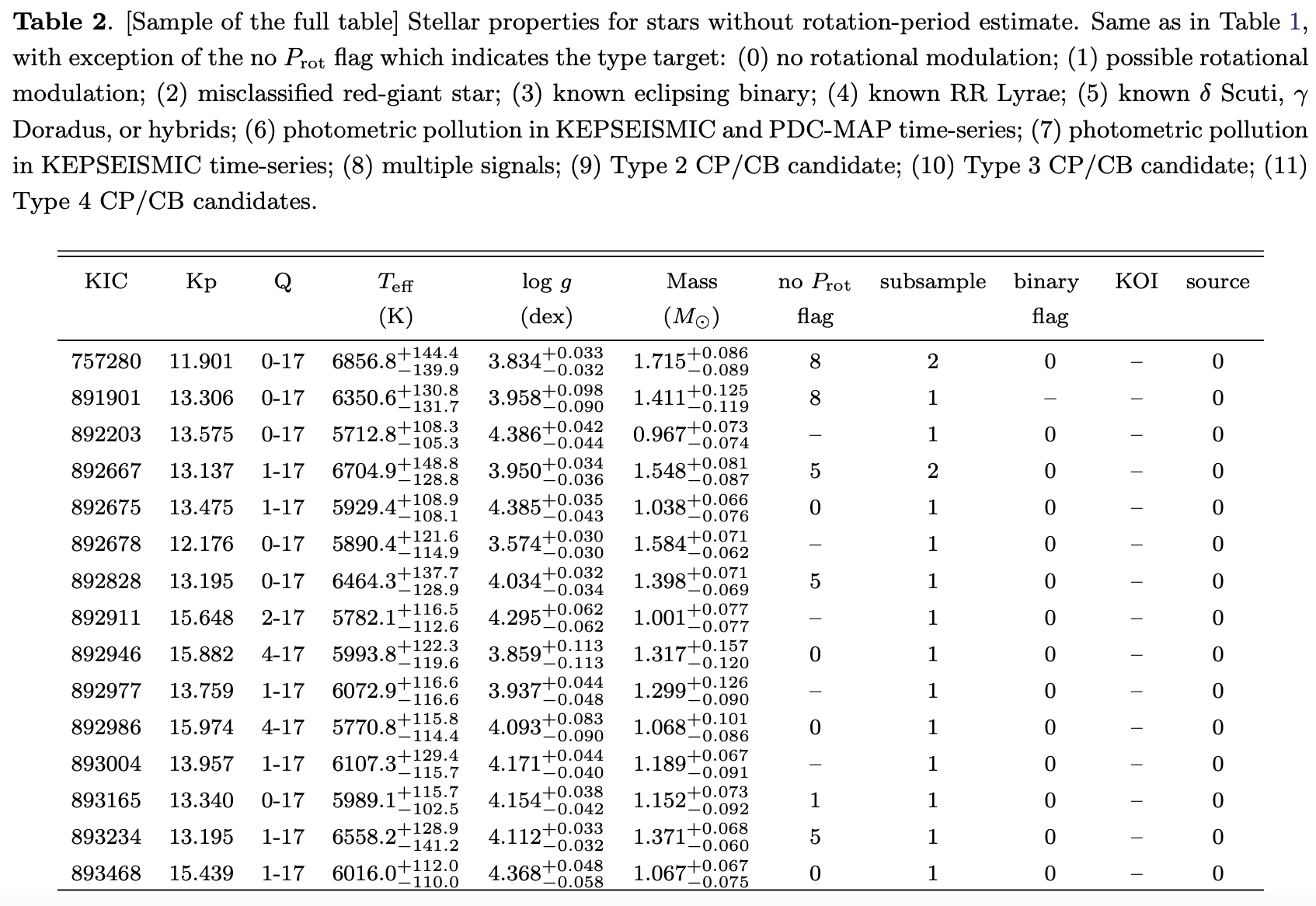}\vspace{0.6cm}
\end{figure*}

\section{Results}\label{sec:results}

Following the methodology described in Sect.~\ref{sec:rotation}, we recover average rotation periods and the respective \sph\ for 39,592 targets from subsamples I-III, which comprise 132,921 targets including part of the training set for \ml. Tables~{\color{blue}1} and {\color{blue}2} summarize the properties of the individual targets with and without \prot\ estimate, respectively. Table~{\color{blue}2} also includes the known contaminants (see Sect.~\ref{sec:sample}) that were not considered in the rotation analysis but are within the parameters space of the subsamples I-III. The final \prot\ yield is summarized in Tables~\ref{tab3} and \ref{tab4}.

\begin{figure}[h]
    \centering
    \includegraphics[width=\hsize]{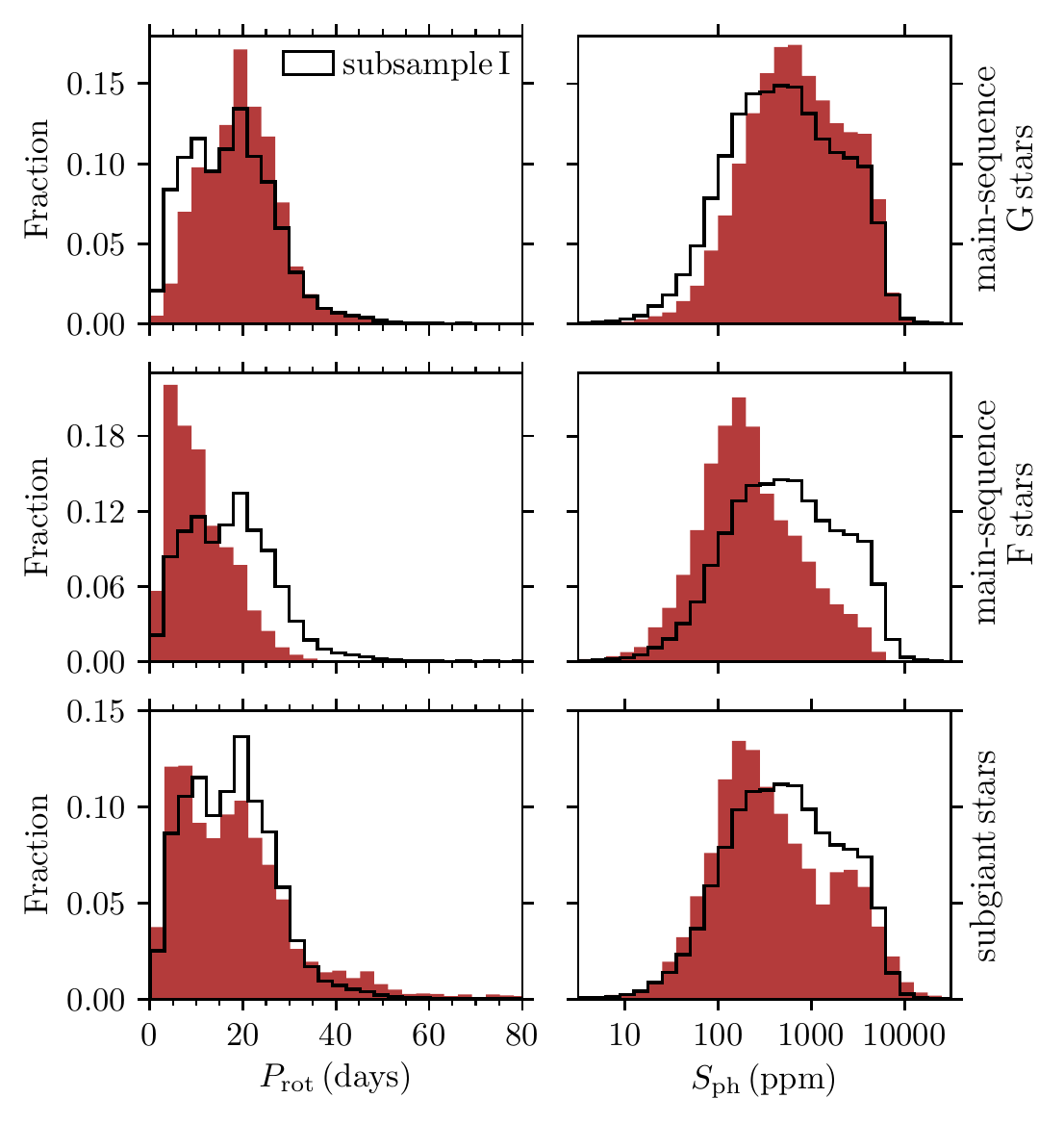}
    \caption{\prot\ (left) and \sph\ (right) distributions for the targets in subsample~I, while neglecting the the \cpcbi\ candidates: full subsample~I (solid black); main-sequence G stars (top row; red); main-sequence F stars (middle row; red); and subgiant stars (bottom row; red).}
    \label{fig:hist}
\end{figure}

\begin{figure}[h]
    \centering
    \includegraphics[width=\hsize]{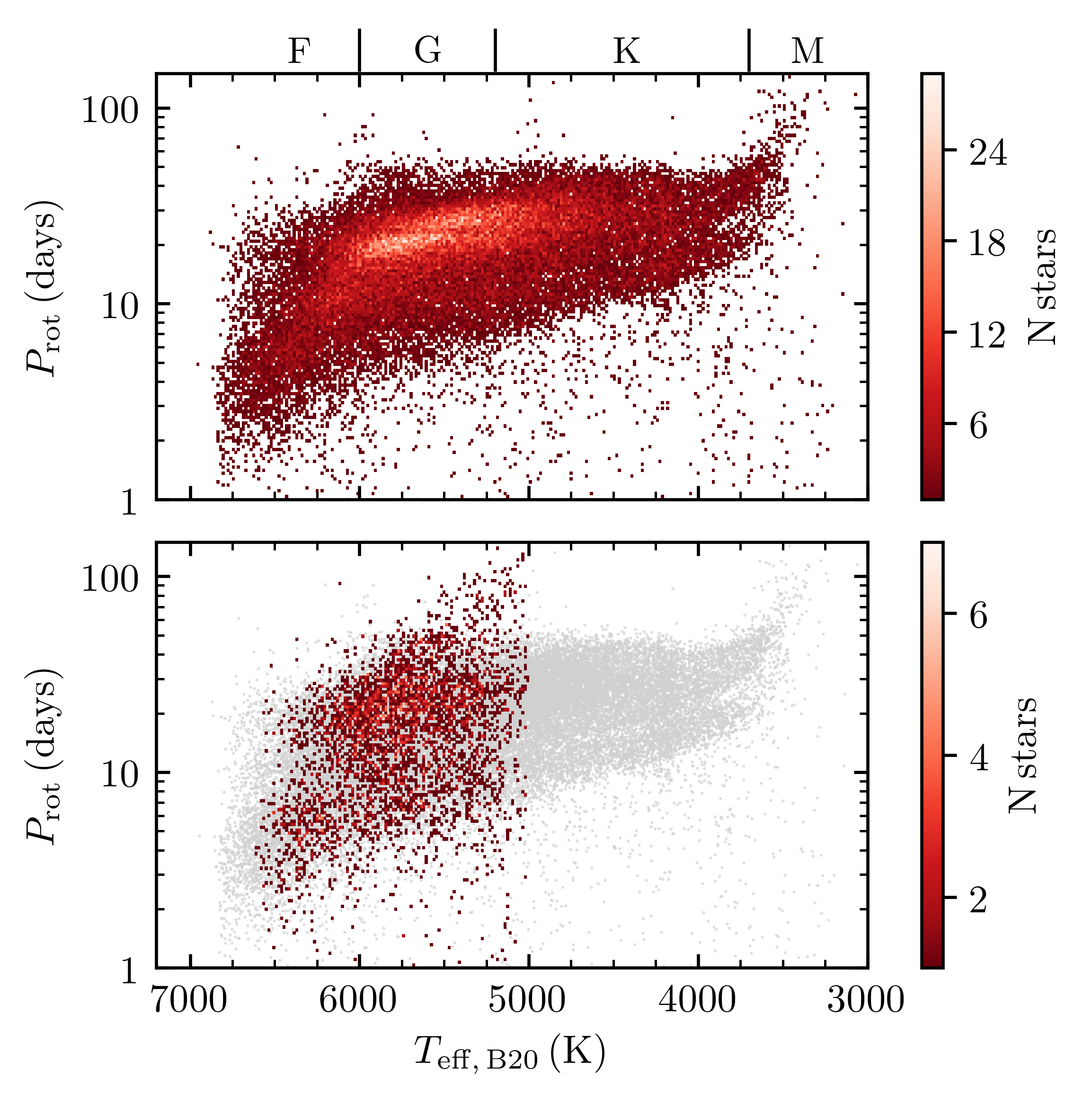}
    \caption{\prot\ as a function of effective temperature for the main-sequence FGKM stars in subsample~I and Paper~I (top) and subgiants (bottom), color-coded by the number of stars. For reference the main-sequence stars in the top panel are represented in gray in the bottom panel. No \cpcbi\ candidates are considered.}
    \label{fig:protteff}
\end{figure}

\begin{figure}[h]
    \centering
    \includegraphics[width=\hsize]{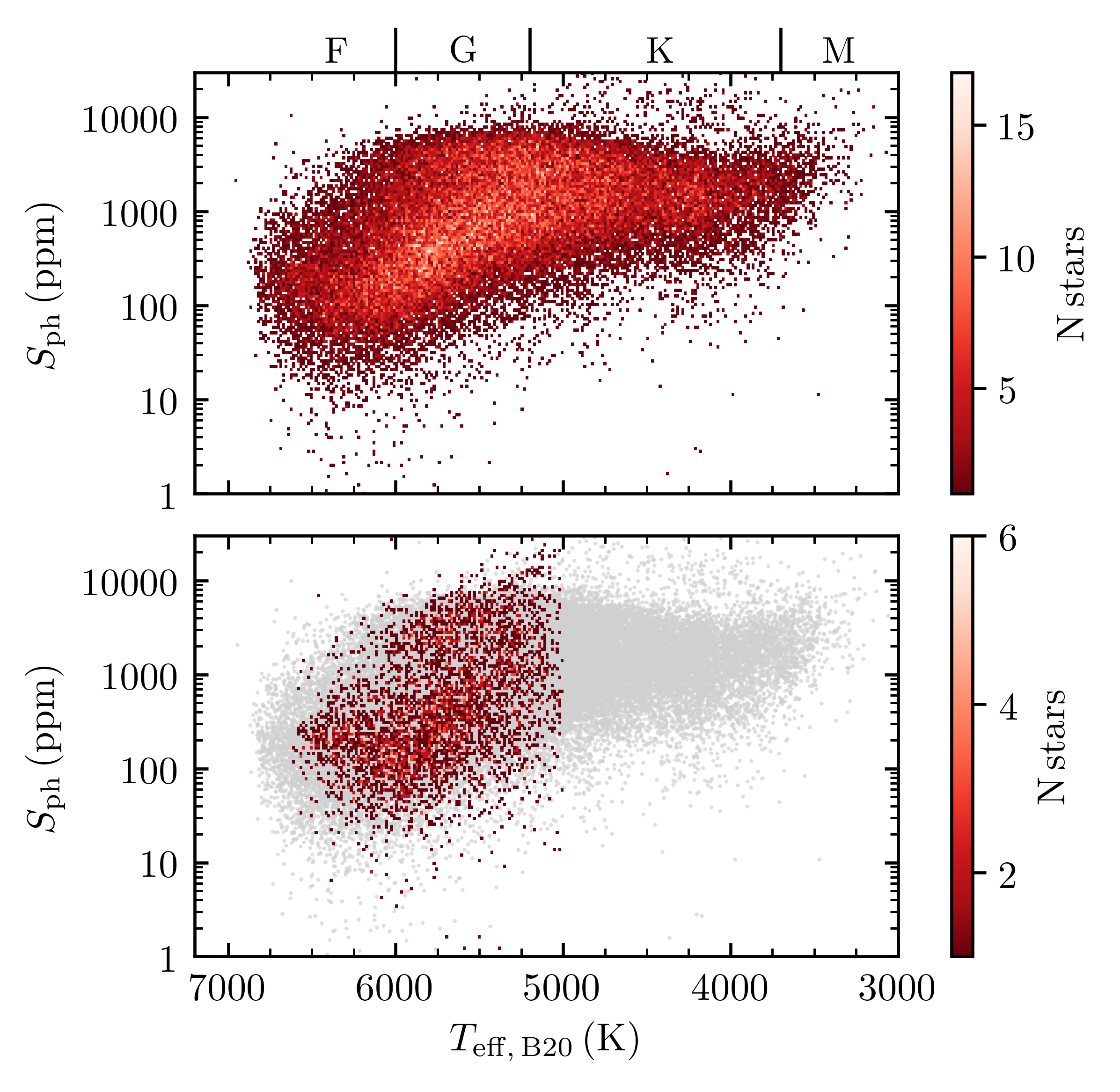}
    \caption{Same as in Fig.~\ref{fig:protteff} but for \sph.}
    \label{fig:sphteff}
\end{figure}

\begin{figure}[h]
    \centering
    \includegraphics[width=\hsize]{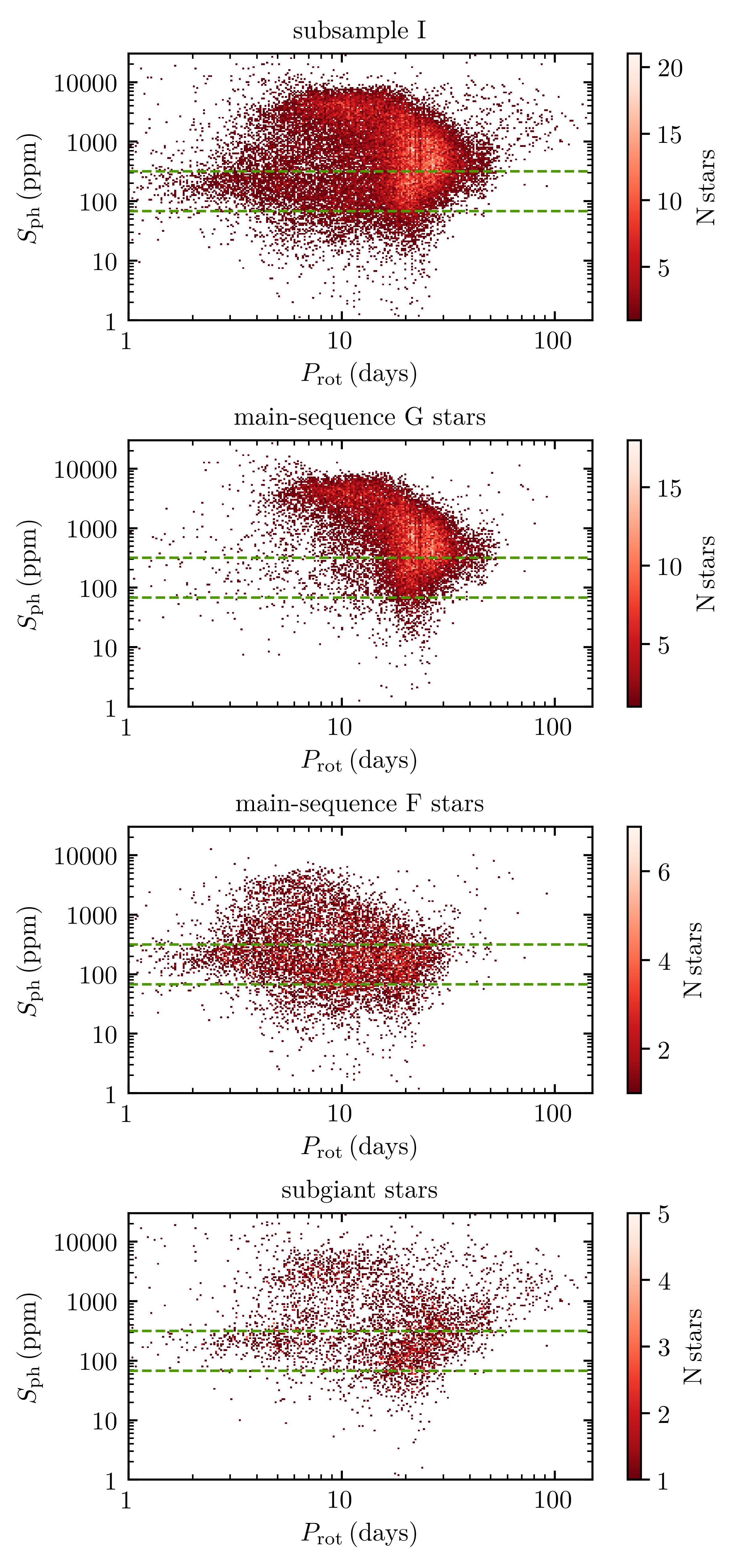}
    \caption{\sph\ as a function of \prot\ for the targets in subsample~I, with exception of the \cpcbi\ candidates, color-coded by number of stars: full subsample~I (top); main-sequence G stars (second row); main-sequence F stars (third row); and subgiant stars (bottom). For reference, the dashed green line marks the solar \sph\ values \citep{Mathur2014} at minimum and maximum of activity.}
    \label{fig:protsph}
\end{figure}

\begin{figure}[h]
    \centering
    \includegraphics[width=\hsize]{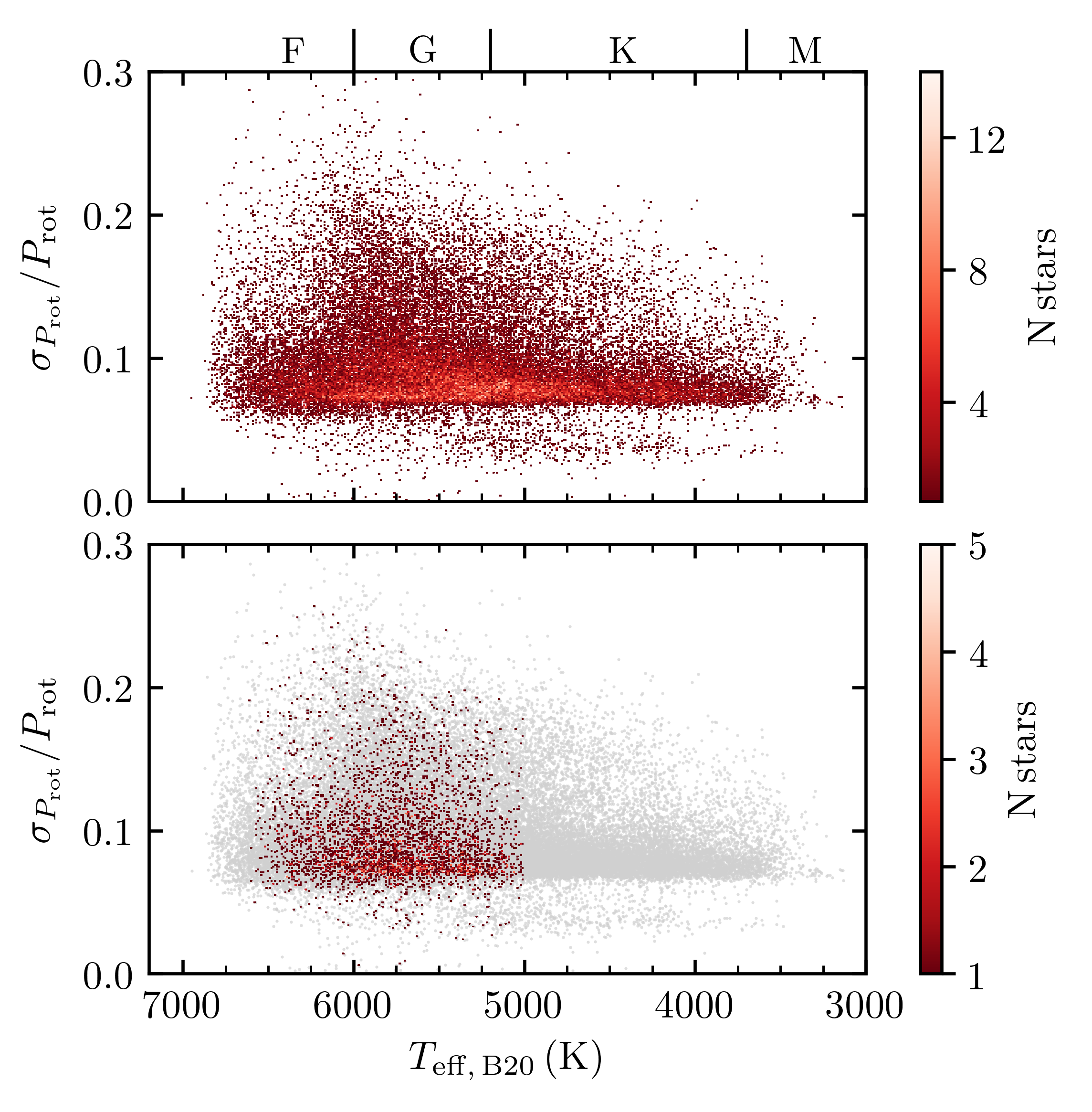}
    \caption{Same as in Fig.~\ref{fig:protteff} but for the relative uncertainty on \prot\ ($\sigma_{P_\text{rot}}/P_\text{rot}$).}
    \label{fig:sigmaprot}
\end{figure}

Figures~\ref{fig:hist}-\ref{fig:sigmaprot} summarize the results for the targets that are solar-like main-sequence and subgiant stars in subsample~I, while neglecting \cpcbi\ candidates flagged either by \ml\ or during the visual inspection. Among subsamples~I-III, we have flagged 2,251 \cpcbi\ candidates.

\begin{table}[h]
\setcounter{table}{2}
\centering\begin{tabular}{rcc}
 & $\#$ \prot &  $\#$ corrected \prot\\
 &  &  ($>15\%$)\\\hline
 
\textsc{ML training set} & &\\
Paper~I & 15,640 & \\
additional targets & 13,923 &  \\
\hline

AutoS & 819 & 2 \\
ML+AutoS & 10,693 & 32 \\
ML & 10,731 & 1099 \\
visual check & 3,426 & \\\hline

total & 55,232 &  \\
new detection & 24,182 (+311) &  \\
& ($\sim43.8\%$) & \\\hline

\end{tabular}
\caption{Summary of the \prot\ detections. {\it Top:} Machine learning (ML) training set composed of the targets in Paper~I and additional targets to complement the \teff\ range. Reliable rotation periods in the training set are obtained by combination of automatic selection (AutoS) and visual inspection. {\it Middle:} Summary of the results for the targets analyzed by \ml\ (see details in Sect.~\ref{sec:vischecks}). The rotation periods here are recovered by automatic selection, machine learning, and additional visual checks. {\it Bottom:} Total number of \prot\ detections and number of new detections in comparison with (McQ14; +311 corresponds to incorrect \prot\ values reported in the literature). First column shows the number of \prot\ detections in each step. The second column indicates the number of \prot\ values that were corrected after visual inspection for the targets in the \ml\ analysis.}\label{tab3}\end{table}

\begin{table}[h]
\centering\begin{tabular}{rrccc}
& & with & without & detection \\
& & \prot & \prot & fraction\\\hline
& \textsc{Paper~I} & 15,640 & 9,415 & 62.4\% \\\hline

\parbox[t]{2mm}{\multirow{6}{*}{\rotatebox[origin=c]{90}{\textsc{This work}}}} 
& \textsc{Subsample I} & & &\\ 
& MS K stars* & 3,275 & 3,153 & 51.0\% \\
& MS G stars & 18,951 & 41,891 & 31.1\% \\
& MS F stars & 9,455 & 22,776 & 29.3\% \\
& subgiant stars & 4,515 & 17,733 & 20.3\% \\\cline{2-5}

& \textsc{Subsample II} & 2,794 & 6,471 & 30.2\% \\ 

& \textsc{Subsample III} & 602 & 1,305 & 31.6\% \\ \hline
\end{tabular}
\caption{Summary of the overall results. The top rows indicate the results from Paper~I (main-sequence KM stars in DR25), while the middle and bottom rows indicate the results from this work. Subsample~I is split according to the stellar properties in B20. *For that reason the K-dwarf sample is not complete as part was already analyzed in Paper~I. Known contaminants are not taken into account here (Sect.~\ref{sec:sample}).}\label{tab4}\end{table}

Figure~\ref{fig:hist} shows the \prot\ and \sph\ distributions per type of targets (main-sequence F and G stars, and subgiants) in comparison with the distributions for the full subsample~I. Note that because of the updated stellar properties, subsample~I contains some K dwarfs according to B20. These are not represented individually in this section (see Appendix~\ref{app:allB20} instead). The dependency on effective temperature is better depicted in Figs.~\ref{fig:protteff} and \ref{fig:sphteff}, which show \prot\ and \sph\ as a function of \teff, color-coded by the number of targets. For reference, Figs.~\ref{fig:protteff} and \ref{fig:sphteff} also include the targets from Paper~I with \prot. Although less pronounced than for the cooler stars, the \prot\ distribution for the hotter stars (subsample~I) also shows evidence for bimodality. The \sph\ distribution tends to be shifted towards smaller \sph\ values for hotter stars than for cooler stars, with F stars showing lower levels of photometric activity than GKM stars. From M to G stars, the range of \sph\ values becomes wider, with the upper and lower edge taking place at larger and smaller \sph\ as \teff\ increases, respectively. As for the subgiant stars, although their \prot\ distribution is similar to the main-sequence stars of similar \teff, there are more slower rotators among subgiants. In particular, between $\sim 5000$ and $\sim 6000$ K there is a group of slow-rotating subgiants, which are relatively cool and evolved subgiants (see Fig.~\ref{fig:hr_subG_slow} in Appendix~\ref{app:subG}). For cooler subgiants, the \sph\ distribution is similar to the main-sequence stars' distribution. However, the hotter subgiants have distinctively low photometric activity levels.

Figure~\ref{fig:protsph} shows the \sph\ as a function of \prot\ for the targets in subsample~I, except for the \cpcbi\ candidates. For main-sequence G stars, fast rotators are typically more photometrically active than slow rotators (Spearman correlation coefficient of -0.45). At relatively short \prot, \sph\ saturates. Part of the main-sequence F stars and subgiants also show the same behavior, but a new group of hot weakly active stars is apparent (Spearman correlation coefficients of -0.16 and -0.07, respectively). In particular, the weakly active fast rotating F stars correspond mainly to targets expected to be above the Kraft break \citep[][see Appendix~\ref{app:Fstars}]{kraft_studies_1967}.

Finally, the relative uncertainty on \prot\ is depicted in Fig.~\ref{fig:sigmaprot}. As described in Sect.~\ref{sec:rotation} we prioritize the rotation-period estimate from the GWPS, where the width of the rotation peak reflects both the uncertainty on the rotation determination and partially differential rotation. The average uncertainty for main-sequence and subgiant solar-like stars (except \cpcbi\ candidates) is about 10\%. The \prot\ uncertainty generally increases with \teff. Interestingly, the maximum relative \prot\ uncertainty is reached around the Kraft break.

Appendix~\ref{app:allB20} shows the same as Figs.~\ref{fig:protteff}-\ref{fig:protsph} but for all targets with \prot\ estimate. In particular, as described in Sect.~\ref{sec:sample}, the \cpcbi\ candidates tend to be fast rotators with large-amplitude brightness variations. Thus, when considering \cpcbi\ candidates, there is an increase of fast rotators with very large \sph, namely values that are larger than the typical \sph\ values for stars of similar \teff. Also, \cpcbi\ candidates often have small \prot\ uncertainties, being in average 7\%.

\subsection{Comparison with the original \ml\ results}

\begin{figure}[h]
    \centering
    \includegraphics[width=\hsize]{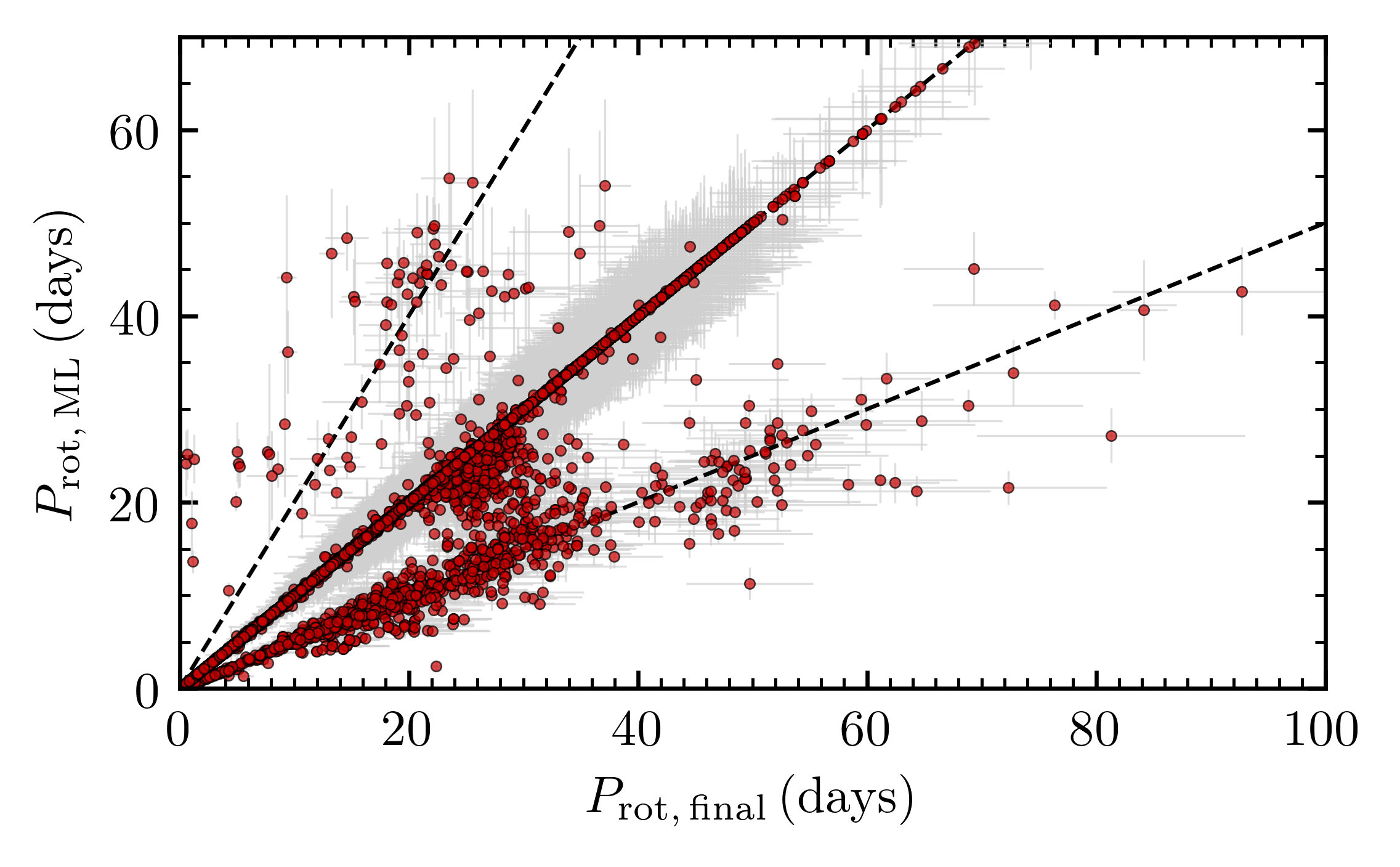}
    \caption{Comparison between \protml\ and the final \prot\ values. The final \prot\ is represented in the horizontal axis for easy comparison with Fig.~6 in \citet{Breton2021}. The dashed lines mark the 1-1, 1-2, and 2-1 lines.}
    \label{fig:final_ML}
\end{figure}

Figure~\ref{fig:final_ML} shows the comparison between the original \protml\ selected by \ml\ and the final \prot\ adopted after the additional visual inspection described in Sect.~\ref{sec:vischecks}. For comparison purposes, in this section and Sect.~\ref{sec:McQ}, the final \prot\ values are indicated by $P_\text{rot,\,final}$ instead of simply \prot. The total number of targets analysed by \ml\ is 98,821. Not accounting with known contaminants (Sect.~\ref{sec:sample}), \ml\ selected 23,547 \protml. Following the steps in Sect.~\ref{sec:vischecks}, 2123 targets were demoted to the Table~{\color{blue}2}, while 21,424 are among the targets with final \prot. From the latter, \protml\ agrees within 15\% with the final \prot\ for 20,293. These results indicate that the global \ml's accuracy is 86.2\%. For the targets in disagreement, 68.1\% of those are related to cases where \protml\ is in fact the second harmonic (one half) of \prot. Another problematic group for \ml\ corresponds to targets with \protml\ between $\sim40$ and $\sim 50$ days \citep[see for example Fig.~4 in][]{Breton2021}. Nevertheless, these account for only a small fraction of the targets. Finally, another group of targets in slight disagreement (still within 15\%)  correspond to targets with final rotation periods around mid-twenties, which reflect the impact from the filtering of the light curve. As described in Sect.~\ref{sec:vischecks}, for part of these targets the rotation period was automatically changed from \protml\ to the final \prot\ value, namely that obtained from the GWPS of the 55-day filtered light curves.

During the visual inspection, \prot\ was recovered for 3,426 additional targets (out of the 98,821) for which \ml\ did not provide a rotation period.

\subsection{Comparison with McQuillan et al. (2014)}\label{sec:McQ}

\begin{figure}[h]
    \centering
    \includegraphics[width=\hsize]{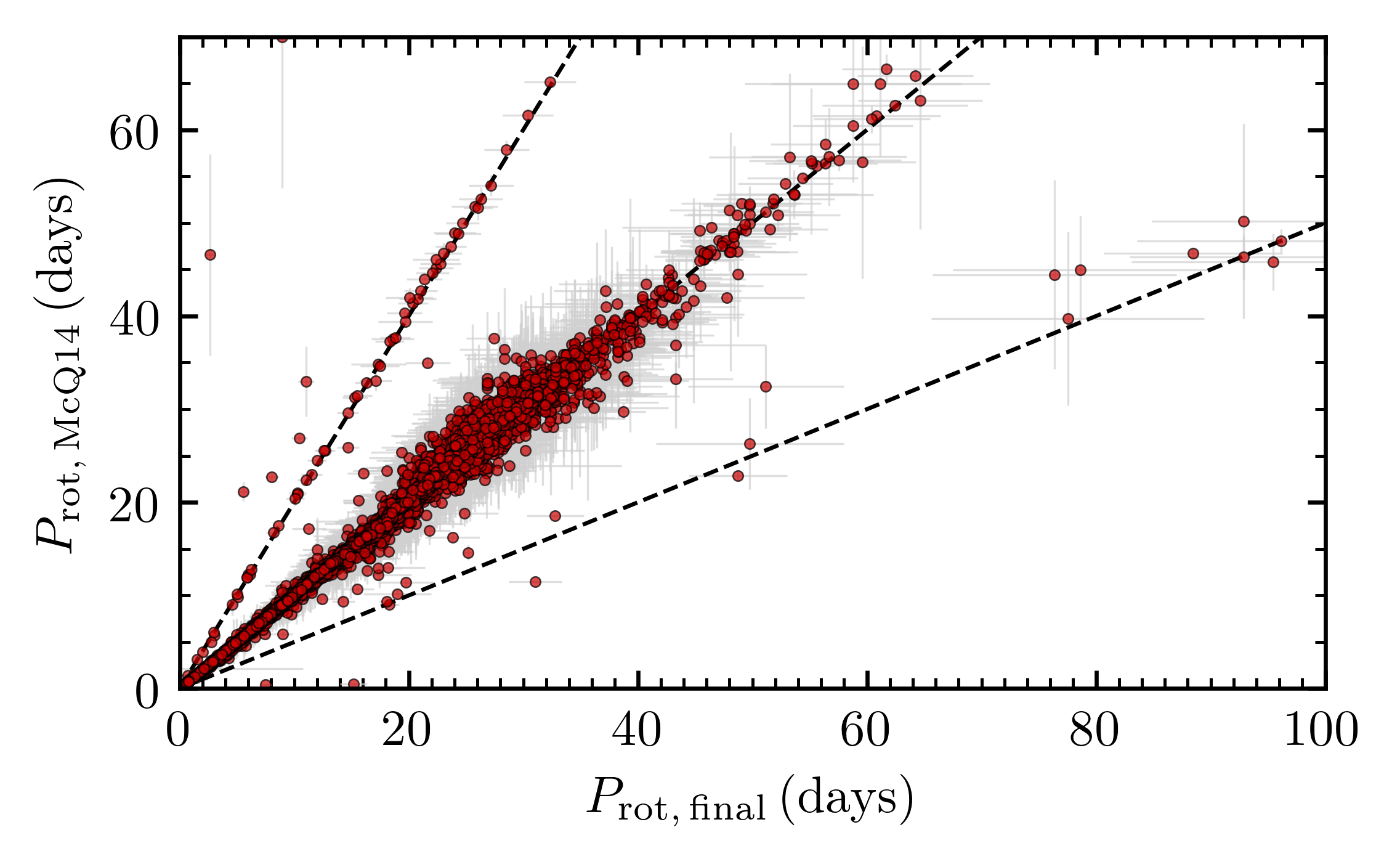}
    \caption{Comparison between the \prot\ values reported by McQ14 and the final \prot\ values reported in this work. The dashed lines mark the 1-1, 1-2, and 2-1 lines.}
    \label{fig:McQ}
\end{figure}

Figure~\ref{fig:McQ} compares the final \prot\ values determined in this work with those reported by McQ14. Among the 20,080 targets in common, there is an agreement within 15\% for 99.1\% of the targets. The \prot\ estimates differ for 183 targets, for which we performed visual checks (Sect.~\ref{sec:vischecks}) and determined that the \prot\ values reported in Table~{\color{blue}1} are correct. Part of the disagreement arises from the fact that the second peak in the ACF can have a larger amplitude than the first, while the first is actually the correct period.

McQ14 reported \prot\ for 615 known contaminants (see details in Sect.~\ref{sec:sample}) within the parameter space of the target sample of this work (i.e. subsamples I-III): 553 red giants; 22 $\delta$ Scuti, $\gamma$ Doradus, or hybrids; 28 eclipsing binaries; 12 RR Lyrae stars. In addition to the known contaminants, McQ14 also reported periods for light curves with photometric pollution or instrumental modulation (for which is not possible to disentangle the intrinsic rotation signal or simply do not show rotational modulation), and Type~2-4 CP/CB candidates. 

Considering the full sample of main-sequence and subgiant FGKM stars (this work and Paper~I), we report \prot\ for 31,038 targets in McQ14, with an agreement (within 15\%) of 99.0\%. Note that the targets for which we do not report \prot\ are now known contaminants or targets for which during the visual inspection we determined that they belong to Table~{\color{blue}2}. We report \prot\ for 24,182 main-sequence and subgiant FGKM stars that were not part of the periodic table of McQ14. 15,088 of those were listed as non-periodic stars in McQ14: the period assessment agrees within 15\% with our final values for 55.5\% of the targets. 3,632 stars (out of 15,088) in the McQ14 non-periodic table do not have a period candidate.

\begin{figure}[h]
    \centering
    \includegraphics[width=\hsize]{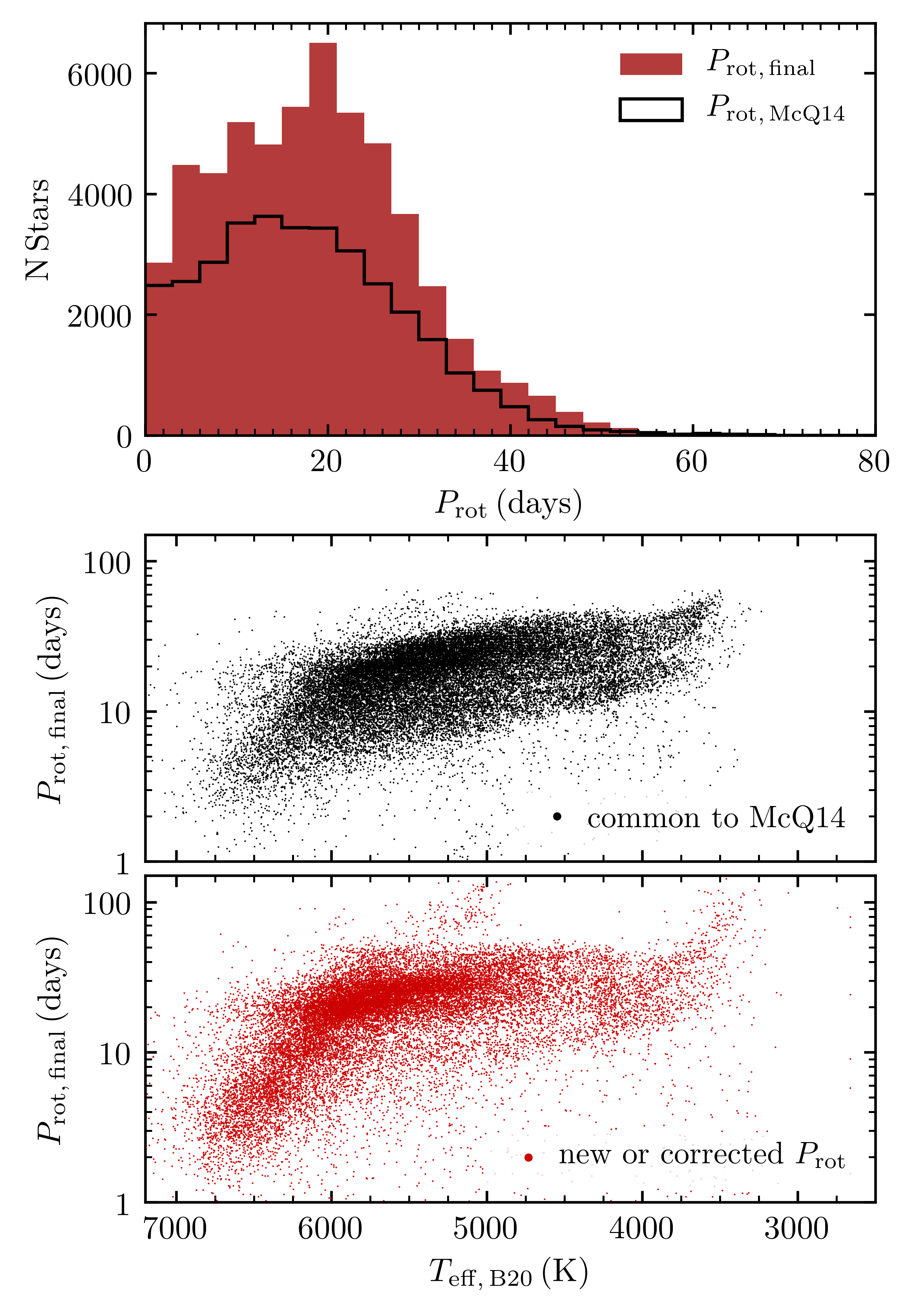}
    \caption{{\it Top:} Comparison between the \prot\ distribution for the targets of the current work and Paper~I (red) and that of McQ14 (black). {\it Middle:} Common \prot\ detections between this work and McQ14 in the \prot-\teff\ diagram. {\it Bottom:} New \prot\ detections in the \prot-\teff\ diagram.}
    \label{fig:histMcQ}
\end{figure}

Figure~\ref{fig:histMcQ} compares the \prot\ distribution for the combined results of Paper~I and the current analysis (red) with that from McQ14 (black). The bottom panels illustrate where the new \prot\ detections lie in the \prot-\teff\ diagram in comparison with those in McQ14. We recover rotation periods for a larger number of fast-rotating F stars and, particularly, for a larger number of GKM slower rotators. While the new \prot\ estimates alter the upper edge of the \prot\ distribution, they do not alter the previous findings on the bimodal \prot\ distribution in the \kep\ field (e.g. McQ14) nor the subsequent gap, i.e. region of low density (see further discussion below).

\begin{figure}[h]
    \centering
    \includegraphics[width=\hsize]{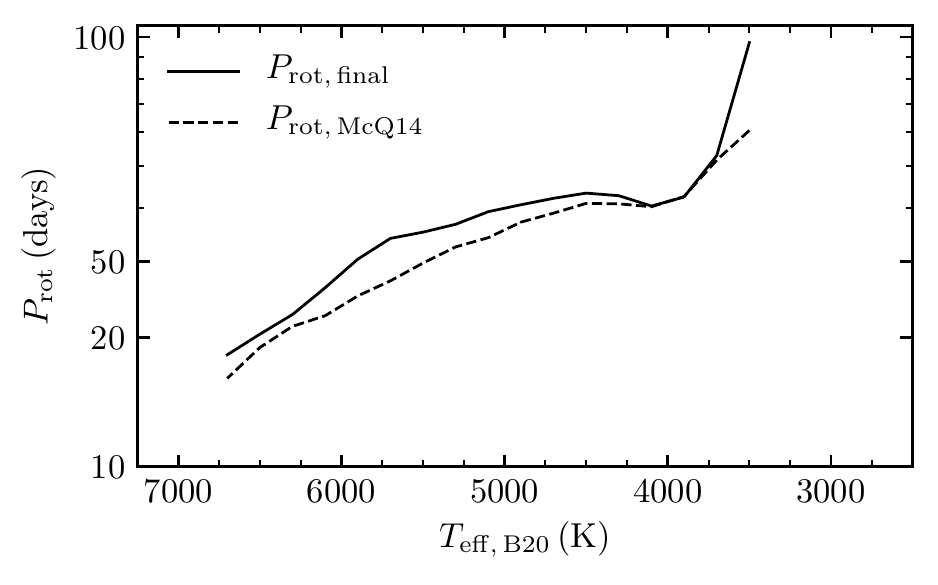}
    \caption{Upper edge (95\% percentile) of the \prot\ distribution found in this work (solid black line) and in McQ14 (dashed black line).}
    \label{fig:envelope}
\end{figure}

Figure~\ref{fig:envelope} shows the upper edge of the \prot\ distribution obtained in this work (solid black line) in comparison with that for the results of McQ14 (dashed black line). The upper edge corresponds to the 95\% percentile. Within $T_\text{eff}\sim 3500$ K and $\sim 4000$ K there is a reasonable agreement between the two upper edges (the same for the DR25 catalog - Appendix~\ref{app:allB20} - except they are shifted towards cooler temperatures). Outside this \teff\ range, we recover rotation periods for a larger number of slow rotators in comparison with McQ14 (see Fig.~\ref{fig:histMcQ}). Therefore, the upper edge recovered in this work is located at longer \prot. This result may be consistent with the model predictions by \citet{vanSaders2019}, which indicates a larger fraction of slow rotators than that detected by McQ14.

\section{Summary and Conclusions}\label{sec:conclusions}

The rotational modulation of light curves due to dark-magnetic spots co-rotating with the stellar surface allows us to constrain rotation and magnetic activity properties. In this work, in order to recover average rotation periods and photometric magnetic activity we analyse the long-cadence data collected by \kep\ for 132,921 stars, that were originally selected according to DR25 \citep{Mathur2017} as main-sequence F and G stars and late subgiant stars. This work is the second of this series, where Paper~I focused the analysis to main-sequence K and M stars (according to DR25).

In this work, we decided to adopt the recent update, using {\it Gaia} data, on the stellar properties for \kep\ targets \citep[B20;][]{Berger2020}. Therefore, some of the targets in Paper~I are now hotter stars (namely G dwarfs), while some targets originally selected for the current work are K dwarfs according to B20.

Our study uses KEPSEISMIC \citep{Garcia2011,Garcia2014a,Pires2015} time-series obtained with three different filters with cutoff periods at 20, 55, and 80 days. The parallel analysis of the three time-series aims at avoiding the long-term instrumental modulations, while retrieving the rotation period, i.e. unaffected by the filtering process. We also use PDC-MAP light curves to determine whether the measured signal could be due to photometric pollution resulting from the larger apertures employed in KEPSEISMIC data.


Rotation-period candidates are retrieved by combining the wavelet analysis with the autocorrelation function of light curves \citep[e.g.][]{Mathur2010,Garcia2014,Ceillier2016,Ceillier2017,Santos2019a}. The final \prot\ estimates are selected by a machine learning algorithm \citep[\ml;][]{Breton2021}, automatic selection, and complementary visual examination. The training set for \ml\ includes the targets of Paper~I \citep{Santos2019a} and 34,100 additional targets analysed in the current work to cover the full \teff\ range. \ml\ then searches for rotational signals and the respective rotation periods among the remaining 98,821 targets. Finally, we perform a series of cross-checks and supplementary visual checks.

We compute the photometric activity proxy as the standard deviation over light curve segments of length $5\times P_\text{rot}$ \citep{Mathur2014}. The final reported \sph\ corresponds to the average of the individual \sph\ values. Although \sph\ is a lower limit of the true photometric activity level, depending for example on the stellar inclination angle and on the longitudinal and latitudinal spot distribution, \sph\ has been shown to be an adequate magnetic activity proxy \citep{Salabert2016a,Salabert2017}.

We report surface rotation periods and the respective \sph \ for 39,592 main-sequence and subgiant solar-like stars (out of 132,921).
In comparison with Paper~I, focused on cooler stars, there is a significant decrease in the detection fraction. The detection fraction in Paper~I was about 60\%, while the detection in this work is about 30\%. A drastic decrease in the detection fraction with \teff\ was also observed by \citet[][McQ14]{McQuillan2014}. In particular, F~stars seem to have rotational modulation with distinct characteristics from those of cooler stars. This motivated the expansion of the training set for \ml\ to properly account for the different behavior of the hottest stars considered here. The change in behavior may be due to the shallow convective zones in F~stars. The amplitude of rotational modulation is distinctively small, which can reflect weak magnetic activity characterized by small, less, and/or short-lived spots or active regions. We also find that the rotational signal of F stars is typically complex, with broad rotation peaks in the GWPS (relatively large \prot\ uncertainties) and multiple peaks in the power spectrum. Often, in the WPS we observe a blended band of stronger rotational signal that ranges from the first harmonic (\prot) to the third harmonic. This is in contrast with the signal of cooler stars (see for example Fig.~\ref{fig:a2ztool}).

2,251 targets (out of 39,592) are flagged as \cpcbi\ candidates as their signal also does not seem to be consistent with that of the other solar-like rotators. These targets generally have short \prot\ and large \sph\ (see Appendix~\ref{app:allB20}), being also characterized by stable fast beating in the light curve and a large number of harmonics associated to \prot. \cpcbi\ candidates tend to be beyond or close to the upper edge of the \sph\ distribution and the lower edge of the \prot\ distribution. Interestingly, in Paper~I, we verified a significant overlap between the \cpcbi\ candidates and the synchronized binaries identified by \citet{Simonian2019}, suggesting thus the possibility of the \cpcbi\ being close-in binaries. 

For the target sample of the current work, we report \prot\ of 19,732 targets for which McQ14 did not report a \prot. For the common targets, there is an agreement of more than 99\%. Note that the majority of the \prot\ values reported here correspond to targets that are expected to be main-sequence solar-like stars. Therefore, even ignoring the detections for subgiant stars which were not the focus in McQ14, our analysis still yields a significantly larger number of \prot\ detections. Nevertheless, McQ14 reported \prot\ for 2,060 targets that are considered subgiants in this work. Note also that the stellar properties (e.g. \teff\ and \logg) have been updated since the study by McQ14.

The rotation period generally decreases with increasing effective temperature, with F stars being on average faster rotators than the cooler solar-like stars. This is consistent with previous findings \citep[e.g.][]{McQuillan2014,Garcia2014}.

Relative to the rotation-period distribution reported by McQ14, we recover a larger number of slow rotators. For this reason the upper edge of the \prot\ distribution is located at longer periods than that in McQ14. Interestingly, the model predictions by \citet{vanSaders2019} were consistent with a lager number of slower rotators than that detected by McQ14.

Similarly to the cooler targets of Paper~I, the bimodal \prot\ distribution is found for the targets of the current work. The bimodality in the \prot\ distribution of \kep\ targets was previously identified and investigated by, for example, \citet{McQuillan2013a,McQuillan2014} and \citet{Davenport2017,Davenport2018}. These studies in particular suggested that the bimodal behavior is related to two distinct episodes of stellar formation. This bimodal behavior is, however, not exclusive to the targets in the \kep\ field and was also discovered for K2 targets \citep[][]{Reinhold2020,Gordon2021}. An alternative origin for the bimodal \prot\ distribution was suggested by \citet{Montet2017} and \citet{Reinhold2019}, who concluded that the targets in the fast-rotating branch are spot-dominated in contrast with the targets in the slow-rotating branch, which are faculae-dominated. \citet{Gordon2021} proposed instead that the bimodal \prot\ distribution is due to a broken spin-down related to the coupling between the stellar rapidly-rotating core and the envelope.

Among the subgiant stars, there is a group of slow-rotating targets with \teff\ between 5000 and 6000 K. These are found to be consistent with more evolved subgiants. In particular, the slowest of these targets ($P_\text{rot}>60$ days) are located close to the red-giant branch.

The \sph\ values for F stars are significantly smaller than those of cooler main-sequence stars. Considering also the targets of Paper~I, for GKM stars, the range of measured \sph\ values is wider: the \sph\ value corresponding to the upper edge of the \sph\ distribution generally increases with \teff, while the \sph\ value of the lower edge decreases with \teff. For main-sequence GKM stars, \sph\ increases with decreasing \prot, which is consistent with fast rotators being more active than slower rotators \citep[e.g.][]{Vaughan1981,Baliunas1983}. While for K stars (see Appendix~\ref{app:allB20} and Paper~I) the bimodal \prot\ distribution is visible in the \sph-\prot\ diagram through two almost parallel branches, for G stars the fast-rotating branch corresponds to a mostly saturated \sph\ regime. Indeed, particularly for K stars, the transition between the two branches causes a discontinuity in the \sph-\prot\ diagram, where the slowest-rotating stars belonging to the fast-rotating branch have smaller \sph\ values than the fastest-rotating stars belonging to the slow-rotating branch. For K2 targets, \citet{Reinhold2020} used the location of this discontinuity or kink to infer the location of the period gap in the \prot-\teff\ diagram. For main-sequence F stars and subgiants, the correlation between \sph\ and \prot\ is significantly reduced. In particular, the hottest targets are found to be weakly active fast rotators.

Finally, the combined output of Paper~I and the current work is average \prot\ and \sph\ values for 55,232 main-sequence and subgiant FGKM stars (out of 159,442). This results include 24,182 new \prot\ detections in comparison with McQ14.

\acknowledgments
The material is based upon work supported by the National Aeronautics and Space Administration (NASA) under Grant No. NNX17AF27G to the Space Science Institute (Boulder, CO USA), which was the recipient of the grant. This Paper~Includes data collected by the \kep\, mission and obtained from the MAST data archive at the Space Telescope Science Institute (STScI). Funding for the \kep\ mission is provided by the NASA Science Mission Directorate. STScI is operated by the Association of Universities for Research in Astronomy, Inc., under NASA contract NAS 5–26555. The authors thank D. Bossini for providing the evolutionary tracks used in this paper to distinguish main-sequence from subgiant stars. ARGS acknowledges the support STFC consolidated grant ST/T000252/1. SNB and RAG acknowledge the support from PLATO and GOLF CNES grants. SM acknowledges support by the Spanish Ministry of Science and Innovation with the Ramon y Cajal fellowship number RYC-2015-17697 and the grant number PID2019-107187GB-I00. This research has made use of the NASA Exoplanet Archive, which is operated by the California Institute of Technology, under contract with the National Aeronautics and Space Administration under the Exoplanet Exploration Program.

\software{KADACS \citep{Garcia2011}, ROOSTER \citep{Breton2021}, NumPy \citep{2020NumPy-Array}, SciPy \citep{2020SciPy-NMeth}, Matplotlib \citep{matplotlib}, Pandas \citep{mckinney-proc-scipy-2010}, Scikit-learn \citep{scikit-learn}.}

\facility{MAST, \kep\ Eclipsing Binary Catalog, Exoplanet Archive}

\pagebreak
\clearpage

\bibliographystyle{aasjournal}
\bibliography{rotation,python}

\begin{thebibliography}{}
\expandafter\ifx\csname natexlab\endcsname\relax\def\natexlab#1{#1}\fi
\providecommand{\url}[1]{\href{#1}{#1}}

\bibitem[{Abdul-Masih {et~al.}(2016)Abdul-Masih, Pr{\v s}a, Conroy, Bloemen,
  Boyajian, Doyle, Johnston, Kostov, Latham, Matijevi{\v c}, Shporer, \&
  Southworth}]{Adbul-Masih2016}
Abdul-Masih, M., Pr{\v s}a, A., Conroy, K., {et~al.} 2016, AJ, 151, 101.
\newblock \url{http://adsabs.harvard.edu/abs/2016AJ....151..101A}

\bibitem[{Aerts {et~al.}(2019)Aerts, Mathis, \& Rogers}]{Aerts2019}
Aerts, C., Mathis, S., \& Rogers, T.~M. 2019, ARA\&A, 57, 35.
\newblock \url{http://adsabs.harvard.edu/abs/2019ARA%26A..57...35A}

\bibitem[{Aigrain {et~al.}(2015)Aigrain, Llama, Ceillier, Chagas, Davenport,
  Garc{\'i}a, Hay, Lanza, McQuillan, Mazeh, de~Medeiros, Nielsen, \&
  Reinhold}]{Aigrain2015}
Aigrain, S., Llama, J., Ceillier, T., {et~al.} 2015, MNRAS, 450, 3211.
\newblock \url{http://adsabs.harvard.edu/abs/2015MNRAS.450.3211A}

\bibitem[{Angus {et~al.}(2015)Angus, Aigrain, Foreman-Mackey, \&
  McQuillan}]{Angus2015}
Angus, R., Aigrain, S., Foreman-Mackey, D., \& McQuillan, A. 2015, MNRAS, 450,
  1787.
\newblock \url{http://adsabs.harvard.edu/abs/2015MNRAS.450.1787A}

\bibitem[{Angus {et~al.}(2020)Angus, Beane, Price-Whelan, Newton, Curtis,
  Berger, van Saders, Kiman, Foreman-Mackey, Lu, Anderson, \&
  Faherty}]{Angus2020}
Angus, R., Beane, A., Price-Whelan, A.~M., {et~al.} 2020, AJ, 160, 90.
\newblock \url{http://adsabs.harvard.edu/abs/2020AJ....160...90A}

\bibitem[{Baglin {et~al.}(2006)Baglin, Auvergne, Barge, Deleuil, Catala,
  Michel, Weiss, \& {COROT Team}}]{Baglin2006a}
Baglin, A., Auvergne, M., Barge, P., {et~al.} 2006, in Proceedings of "The
  CoRoT Mission Pre-Launch Status - Stellar Seismology and Planet Finding" (ESA
  SP-1306). Editors: M. Fridlund, A. Baglin, J. Lochard and L. Conroy. ISBN
  92-9092-465-9., Vol. 1306, 33.
\newblock \url{http://adsabs.harvard.edu/abs/2006ESASP1306...33B}

\bibitem[{Baliunas {et~al.}(1983)Baliunas, Hartmann, Noyes, Vaughan, Preston,
  Frazer, Lanning, Middelkoop, \& Mihalas}]{Baliunas1983}
Baliunas, S.~L., Hartmann, L., Noyes, R.~W., {et~al.} 1983, ApJ, 275, 752.
\newblock \url{http://adsabs.harvard.edu/abs/1983ApJ...275..752B}

\bibitem[{{Barnes}(2003)}]{Barnes2003}
{Barnes}, S.~A. 2003, \apj, 586, 464.
\newblock \url{https://ui.adsabs.harvard.edu/abs/2003ApJ...586..464B}

\bibitem[{Barnes(2007)}]{Barnes2007}
Barnes, S.~A. 2007, ApJ, 669, 1167.
\newblock \url{http://adsabs.harvard.edu/abs/2007ApJ...669.1167B}

\bibitem[{Beck {et~al.}(2012)Beck, Montalban, Kallinger, De~Ridder, Aerts,
  Garc{\'i}a, Hekker, Dupret, Mosser, Eggenberger, Stello, Elsworth, Frandsen,
  Carrier, Hillen, Gruberbauer, Christensen-Dalsgaard, Miglio, Valentini,
  Bedding, Kjeldsen, Girouard, Hall, \& Ibrahim}]{Beck2012}
Beck, P.~G., Montalban, J., Kallinger, T., {et~al.} 2012, Nature, 481, 55.
\newblock \url{https://www.nature.com/articles/nature10612}

\bibitem[{Benk{\H o} {et~al.}(2010)Benk{\H o}, Kolenberg, Szab{\'o}, Kurtz,
  Bryson, Bregman, Still, Smolec, Nuspl, Nemec, Moskalik, Kopacki, Koll{\'a}th,
  Guggenberger, di~Criscienzo, Christensen-Dalsgaard, Kjeldsen, Borucki, Koch,
  Jenkins, \& van Cleve}]{Benko2010}
Benk{\H o}, J.~M., Kolenberg, K., Szab{\'o}, R., {et~al.} 2010, MNRAS, 409,
  1585.
\newblock \url{http://adsabs.harvard.edu/abs/2010MNRAS.409.1585B}

\bibitem[{Benomar {et~al.}(2015)Benomar, Takata, Shibahashi, Ceillier, \&
  Garc{\'i}a}]{Benomar2015}
Benomar, O., Takata, M., Shibahashi, H., Ceillier, T., \& Garc{\'i}a, R.~A.
  2015, MNRAS, 452, 2654.
\newblock \url{http://adsabs.harvard.edu/abs/2015MNRAS.452.2654B}

\bibitem[{Benomar {et~al.}(2018)Benomar, Bazot, Nielsen, Gizon, Sekii, Takata,
  Hotta, Hanasoge, Sreenivasan, \& Christensen-Dalsgaard}]{Benomar2018}
Benomar, O., Bazot, M., Nielsen, M.~B., {et~al.} 2018, Science, 361, 1231.
\newblock \url{http://adsabs.harvard.edu/abs/2018Sci...361.1231B}

\bibitem[{Berger {et~al.}(2018)Berger, Huber, Gaidos, \& van
  Saders}]{Berger2018}
Berger, T.~A., Huber, D., Gaidos, E., \& van Saders, J.~L. 2018, ApJ, 866, 99.
\newblock \url{http://adsabs.harvard.edu/abs/2018ApJ...866...99B}

\bibitem[{Berger {et~al.}(2020)Berger, Huber, van Saders, Gaidos, Tayar, \&
  Kraus}]{Berger2020}
Berger, T.~A., Huber, D., van Saders, J.~L., {et~al.} 2020, AJ, 159, 280.
\newblock \url{http://adsabs.harvard.edu/abs/2020AJ....159..280B}

\bibitem[{Borucki {et~al.}(2010)Borucki, Koch, Basri, Batalha, Brown, Caldwell,
  Caldwell, Christensen-Dalsgaard, Cochran, DeVore, Dunham, Dupree, Gautier,
  Geary, Gilliland, Gould, Howell, Jenkins, Kondo, Latham, Marcy, Meibom,
  Kjeldsen, Lissauer, Monet, Morrison, Sasselov, Tarter, Boss, Brownlee, Owen,
  Buzasi, Charbonneau, Doyle, Fortney, Ford, Holman, Seager, Steffen, Welsh,
  Rowe, Anderson, Buchhave, Ciardi, Walkowicz, Sherry, Horch, Isaacson,
  Everett, Fischer, Torres, Johnson, Endl, MacQueen, Bryson, Dotson, Haas,
  Kolodziejczak, Van~Cleve, Chandrasekaran, Twicken, Quintana, Clarke, Allen,
  Li, Wu, Tenenbaum, Verner, Bruhweiler, Barnes, \& Prsa}]{Borucki2010}
Borucki, W.~J., Koch, D., Basri, G., {et~al.} 2010, Science, 327, 977.
\newblock \url{http://adsabs.harvard.edu/abs/2010Sci...327..977B}

\bibitem[{Bowman \& Kurtz(2018)}]{Bowman2018}
Bowman, D.~M., \& Kurtz, D.~W. 2018, MNRAS, 476, 3169.
\newblock \url{http://adsabs.harvard.edu/abs/2018MNRAS.476.3169B}

\bibitem[{Bradley {et~al.}(2015)Bradley, Guzik, Miles, Uytterhoeven,
  Jackiewicz, \& Kinemuchi}]{Bradley2015}
Bradley, P.~A., Guzik, J.~A., Miles, L.~F., {et~al.} 2015, AJ, 149, 68.
\newblock \url{https://ui.adsabs.harvard.edu/abs/2015AJ....149...68B/abstract}

\bibitem[{Breton {et~al.}(2021)Breton, Santos, Bugnet, Mathur, Garc{\'i}a, \&
  Pall{\'e}}]{Breton2021}
Breton, S.~N., Santos, A. R.~G., Bugnet, L., {et~al.} 2021, A\&A, 647, A125.
\newblock
  \url{https://www.aanda.org/articles/aa/abs/2021/03/aa39947-20/aa39947-20.html}

\bibitem[{Bugnet {et~al.}(2018)Bugnet, Garc{\'i}a, Davies, Mathur, Corsaro,
  Hall, \& Rendle}]{Bugnet2018}
Bugnet, L., Garc{\'i}a, R.~A., Davies, G.~R., {et~al.} 2018, A\&A, 620, A38.
\newblock \url{http://adsabs.harvard.edu/abs/2018A%26A...620A..38B}

\bibitem[{Ceillier {et~al.}(2016)Ceillier, van Saders, Garc{\'i}a, Metcalfe,
  Creevey, Mathis, Mathur, Pinsonneault, Salabert, \& Tayar}]{Ceillier2016}
Ceillier, T., van Saders, J., Garc{\'i}a, R.~A., {et~al.} 2016, MNRAS, 456,
  119.
\newblock \url{http://adsabs.harvard.edu/abs/2016MNRAS.456..119C}

\bibitem[{Ceillier {et~al.}(2017)Ceillier, Tayar, Mathur, Salabert, Garc{\'i}a,
  Stello, Pinsonneault, van Saders, Beck, \& Bloemen}]{Ceillier2017}
Ceillier, T., Tayar, J., Mathur, S., {et~al.} 2017, A\&A, 605, A111.
\newblock \url{http://adsabs.harvard.edu/abs/2017A%26A...605A.111C}

\bibitem[{Colman {et~al.}(2017)Colman, Huber, Bedding, Kuszlewicz, Yu, Beck,
  Elsworth, Garc{\'i}a, Kawaler, Mathur, Stello, \& White}]{Colman2017}
Colman, I.~L., Huber, D., Bedding, T.~R., {et~al.} 2017, MNRAS, 469, 3802.
\newblock \url{https://ui.adsabs.harvard.edu/abs/2017MNRAS.469.3802C/abstract}

\bibitem[{Davenport(2017)}]{Davenport2017}
Davenport, J. R.~A. 2017, ApJ, 835, 16.
\newblock \url{http://adsabs.harvard.edu/abs/2017ApJ...835...16D}

\bibitem[{Davenport \& Covey(2018)}]{Davenport2018}
Davenport, J. R.~A., \& Covey, K.~R. 2018, ApJ, 868, 151.
\newblock \url{http://adsabs.harvard.edu/abs/2018ApJ...868..151D}

\bibitem[{Deheuvels {et~al.}(2012)Deheuvels, Garc{\'i}a, Chaplin, Basu, Antia,
  Appourchaux, Benomar, Davies, Elsworth, Gizon, Goupil, Reese, Regulo, Schou,
  Stahn, Casagrande, Christensen-Dalsgaard, Fischer, Hekker, Kjeldsen, Mathur,
  Mosser, Pinsonneault, Valenti, Christiansen, Kinemuchi, \&
  Mullally}]{Deheuvels2012}
Deheuvels, S., Garc{\'i}a, R.~A., Chaplin, W.~J., {et~al.} 2012, ApJ, 756, 19.
\newblock \url{http://adsabs.harvard.edu/abs/2012ApJ...756...19D}

\bibitem[{Deheuvels {et~al.}(2014)Deheuvels, Do{\u g}an, Goupil, Appourchaux,
  Benomar, Bruntt, Campante, Casagrande, Ceillier, Davies, De~Cat, Fu,
  Garc{\'i}a, Lobel, Mosser, Reese, Regulo, Schou, Stahn, Thygesen, Yang,
  Chaplin, Christensen-Dalsgaard, Eggenberger, Gizon, Mathis,
  Molenda-{\.Z}akowicz, \& Pinsonneault}]{Deheuvels2014}
Deheuvels, S., Do{\u g}an, G., Goupil, M.~J., {et~al.} 2014, A\&A, 564, A27.
\newblock \url{http://adsabs.harvard.edu/abs/2014A%26A...564A..27D}

\bibitem[{Dupret {et~al.}(2005)Dupret, Grigahc{\`e}ne, Garrido, Gabriel, \&
  Scuflaire}]{Dupret2005}
Dupret, M.-A., Grigahc{\`e}ne, A., Garrido, R., Gabriel, M., \& Scuflaire, R.
  2005, A\&A, 435, 927.
\newblock \url{http://adsabs.harvard.edu/abs/2005A%26A...435..927D}

\bibitem[{Garc{\'i}a \& Ballot(2019)}]{Garcia2019}
Garc{\'i}a, R.~A., \& Ballot, J. 2019, Living Rev. Solar Phys., 16, 4.
\newblock \url{https://ui.adsabs.harvard.edu/2019LRSP...16....4G/abstract}

\bibitem[{Garc{\'i}a {et~al.}(2010)Garc{\'i}a, Mathur, Salabert, Ballot,
  Regulo, Metcalfe, \& Baglin}]{Garcia2010}
Garc{\'i}a, R.~A., Mathur, S., Salabert, D., {et~al.} 2010, Science, 329, 1032.
\newblock \url{http://adsabs.harvard.edu/abs/2010Sci...329.1032G}

\bibitem[{Garc{\'i}a {et~al.}(2011)Garc{\'i}a, Hekker, Stello,
  Guti{\'e}rrez-Soto, Handberg, Huber, Karoff, Uytterhoeven, Appourchaux,
  Chaplin, Elsworth, Mathur, Ballot, Christensen-Dalsgaard, Gilliland, Houdek,
  Jenkins, Kjeldsen, McCauliff, Metcalfe, Middour, Molenda-Zakowicz, Monteiro,
  Smith, \& Thompson}]{Garcia2011}
Garc{\'i}a, R.~A., Hekker, S., Stello, D., {et~al.} 2011, MNRAS, 414, L6.
\newblock \url{http://adsabs.harvard.edu/abs/2011MNRAS.414L...6G}

\bibitem[{Garc{\'i}a {et~al.}(2014{\natexlab{a}})Garc{\'i}a, Ceillier,
  Salabert, Mathur, van Saders, Pinsonneault, Ballot, Beck, Bloemen, Campante,
  Davies, do~Nascimento, Mathis, Metcalfe, Nielsen, Su{\'a}rez, Chaplin,
  Jim{\'e}nez, \& Karoff}]{Garcia2014}
Garc{\'i}a, R.~A., Ceillier, T., Salabert, D., {et~al.} 2014{\natexlab{a}},
  A\&A, 572, A34.
\newblock \url{http://adsabs.harvard.edu/abs/2014A%26A...572A..34G}

\bibitem[{Garc{\'i}a {et~al.}(2014{\natexlab{b}})Garc{\'i}a, Mathur, Pires,
  R{\'e}gulo, Bellamy, Pall{\'e}, Ballot, Barcel{\'o}~Forteza, Beck, Bedding,
  Ceillier, Roca~Cort{\'e}s, Salabert, \& Stello}]{Garcia2014a}
Garc{\'i}a, R.~A., Mathur, S., Pires, S., {et~al.} 2014{\natexlab{b}}, A\&A,
  568, A10.
\newblock \url{http://adsabs.harvard.edu/abs/2014A%26A...568A..10G}

\bibitem[{Gehan {et~al.}(2018)Gehan, Mosser, Michel, Samadi, \&
  Kallinger}]{Gehan2018}
Gehan, C., Mosser, B., Michel, E., Samadi, R., \& Kallinger, T. 2018, A\&A,
  616, A24.
\newblock \url{http://adsabs.harvard.edu/abs/2018A%26A...616A..24G}

\bibitem[{Gordon {et~al.}(2021)Gordon, Davenport, Angus, Foreman-Mackey, Agol,
  Covey, Ag{\"u}eros, \& Kipping}]{Gordon2021}
Gordon, T.~A., Davenport, J. R.~A., Angus, R., {et~al.} 2021, ApJ, 913, 70.
\newblock \url{https://ui.adsabs.harvard.edu/abs/2021ApJ...913...70G}

\bibitem[{Guo {et~al.}(2020)Guo, Shporer, Hambleton, \& Isaacson}]{Guo2020}
Guo, Z., Shporer, A., Hambleton, K., \& Isaacson, H. 2020, ApJ, 888, 95.
\newblock \url{http://adsabs.harvard.edu/abs/2020ApJ...888...95G}

\bibitem[{Hall {et~al.}(2021)Hall, Davies, van Saders, Nielsen, Lund, Chaplin,
  Garc{\'i}a, Amard, Breimann, Khan, See, \& Tayar}]{Hall2021}
Hall, O.~J., Davies, G.~R., van Saders, J., {et~al.} 2021, arXiv:2104.10919
  [astro-ph], doi:10.1038/s41550-021-01335-x.
\newblock \url{http://arxiv.org/abs/2104.10919}

\bibitem[{Harris {et~al.}(2020)Harris, Millman, van~der Walt, Gommers,
  Virtanen, Cournapeau, Wieser, Taylor, Berg, Smith, Kern, Picus, Hoyer, van
  Kerkwijk, Brett, Haldane, Fernández~del Río, Wiebe, Peterson,
  Gérard-Marchant, Sheppard, Reddy, Weckesser, Abbasi, Gohlke, \&
  Oliphant}]{2020NumPy-Array}
Harris, C.~R., Millman, K.~J., van~der Walt, S.~J., {et~al.} 2020, Nature, 585,
  357–362

\bibitem[{Howell {et~al.}(2014)Howell, Sobeck, Haas, Still, Barclay, Mullally,
  Troeltzsch, Aigrain, Bryson, Caldwell, Chaplin, Cochran, Huber, Marcy,
  Miglio, Najita, Smith, Twicken, \& Fortney}]{Howell2014}
Howell, S.~B., Sobeck, C., Haas, M., {et~al.} 2014, PASP, 126, 398.
\newblock \url{http://adsabs.harvard.edu/abs/2014PASP..126..398H}

\bibitem[{Hunter(2007)}]{matplotlib}
Hunter, J.~D. 2007, Computing in Science \& Engineering, 9, 90.
\newblock \url{https://doi.org/10.1109/MCSE.2007.55}

\bibitem[{Jenkins {et~al.}(2010)Jenkins, Caldwell, Chandrasekaran, Twicken,
  Bryson, Quintana, Clarke, Li, Allen, Tenenbaum, Wu, Klaus, Middour, Cote,
  McCauliff, Girouard, Gunter, Wohler, Sommers, Hall, Uddin, Wu, Bhavsar,
  Van~Cleve, Pletcher, Dotson, Haas, Gilliland, Koch, \& Borucki}]{Jenkins2010}
Jenkins, J.~M., Caldwell, D.~A., Chandrasekaran, H., {et~al.} 2010, ApJL, 713,
  L87.
\newblock \url{http://adsabs.harvard.edu/abs/2010ApJ...713L..87J}

\bibitem[{Kirk {et~al.}(2016)Kirk, Conroy, Pr{\v s}a, Abdul-Masih, Kochoska,
  Matijevi{\v c}, Hambleton, Barclay, Bloemen, Boyajian, Doyle, Fulton,
  Hoekstra, Jek, Kane, Kostov, Latham, Mazeh, Orosz, Pepper, Quarles,
  Ragozzine, Shporer, Southworth, Stassun, Thompson, Welsh, Agol, Derekas,
  Devor, Fischer, Green, Gropp, Jacobs, Johnston, LaCourse, Saetre,
  Schwengeler, Toczyski, Werner, Garrett, Gore, Martinez, Spitzer, Stevick,
  Thomadis, Vrijmoet, Yenawine, Batalha, \& Borucki}]{Kirk2016}
Kirk, B., Conroy, K., Pr{\v s}a, A., {et~al.} 2016, AJ, 151, 68.
\newblock \url{http://adsabs.harvard.edu/abs/2016AJ....151...68K}

\bibitem[{Kraft(1967)}]{kraft_studies_1967}
Kraft, R.~P. 1967, APJ, 150, 551.
\newblock \url{http://adsabs.harvard.edu/abs/1967ApJ...150..551K}

\bibitem[{Lee {et~al.}(2016)Lee, Hong, Koo, \& Park}]{Lee2016}
Lee, J.~W., Hong, K., Koo, J.-R., \& Park, J.-H. 2016, ApJ, 820, 1.
\newblock \url{https://doi.org/10.3847%2F0004-637x%2F820%2F1%2F1}

\bibitem[{Li {et~al.}(2019{\natexlab{a}})Li, Bedding, Murphy, Van~Reeth,
  Antoci, \& Ouazzani}]{Li2019b}
Li, G., Bedding, T.~R., Murphy, S.~J., {et~al.} 2019{\natexlab{a}}, MNRAS, 482,
  1757.
\newblock \url{http://adsabs.harvard.edu/abs/2019MNRAS.482.1757L}

\bibitem[{Li {et~al.}(2019{\natexlab{b}})Li, Van~Reeth, Bedding, Murphy, \&
  Antoci}]{Li2019}
Li, G., Van~Reeth, T., Bedding, T.~R., Murphy, S.~J., \& Antoci, V.
  2019{\natexlab{b}}, MNRAS, 487, 782.
\newblock \url{http://adsabs.harvard.edu/abs/2019MNRAS.487..782L}

\bibitem[{Liu {et~al.}(2007)Liu, San~Liang, \& Weisberg}]{Liu2007}
Liu, Y., San~Liang, X., \& Weisberg, R.~H. 2007, J. Atmos. Oceanic Technol.,
  24, 2093.
\newblock \url{https://journals.ametsoc.org/doi/abs/10.1175/2007JTECHO511.1}

\bibitem[{Lorenzo-Oliveira {et~al.}(2020)Lorenzo-Oliveira, Mel{\'e}ndez, Ponte,
  \& Galarza}]{LorenzoOliveira2020}
Lorenzo-Oliveira, D., Mel{\'e}ndez, J., Ponte, G., \& Galarza, J.~Y. 2020,
  Monthly Notices of the Royal Astronomical Society, 495, L61.
\newblock \url{http://adsabs.harvard.edu/abs/2020MNRAS.495L..61L}

\bibitem[{Lorenzo-Oliveira {et~al.}(2019)Lorenzo-Oliveira, Mel{\'e}ndez,
  Yana~Galarza, Ponte, dos Santos, Spina, Bedell, Ram{\'i}rez, Bean, \&
  Asplund}]{Lorenzo-Oliveira2019}
Lorenzo-Oliveira, D., Mel{\'e}ndez, J., Yana~Galarza, J., {et~al.} 2019, MNRAS,
  485, L68.
\newblock \url{http://adsabs.harvard.edu/abs/2019MNRAS.485L..68L}

\bibitem[{Mamajek \& Hillenbrand(2008)}]{Mamajek2008}
Mamajek, E.~E., \& Hillenbrand, L.~A. 2008, ApJ, 687, 1264.
\newblock \url{http://adsabs.harvard.edu/abs/2008ApJ...687.1264M}

\bibitem[{Mathur {et~al.}(2010)Mathur, Garc{\'i}a, Catala, Bruntt, Mosser,
  Appourchaux, Ballot, Creevey, Gaulme, Hekker, Huber, Karoff, Piau,
  R{\'e}gulo, Roxburgh, Salabert, Verner, Auvergne, Baglin, Chaplin, Elsworth,
  Michel, Samadi, Sato, \& Stello}]{Mathur2010}
Mathur, S., Garc{\'i}a, R.~A., Catala, C., {et~al.} 2010, A\&A, 518, A53.
\newblock \url{http://adsabs.harvard.edu/abs/2010A%26A...518A..53M}

\bibitem[{Mathur {et~al.}(2014)Mathur, Garc{\'i}a, Ballot, Ceillier, Salabert,
  Metcalfe, R{\'e}gulo, Jim{\'e}nez, \& Bloemen}]{Mathur2014}
Mathur, S., Garc{\'i}a, R.~A., Ballot, J., {et~al.} 2014, A\&A, 562, A124.
\newblock \url{http://adsabs.harvard.edu/abs/2014A%26A...562A.124M}

\bibitem[{Mathur {et~al.}(2017)Mathur, Huber, Batalha, Ciardi, Bastien,
  Bieryla, Buchhave, Cochran, Endl, Esquerdo, Furlan, Howard, Howell, Isaacson,
  Latham, MacQueen, \& Silva}]{Mathur2017}
Mathur, S., Huber, D., Batalha, N.~M., {et~al.} 2017, ApJS, 229, 30.
\newblock \url{http://adsabs.harvard.edu/abs/2017ApJS..229...30M}

\bibitem[{McQuillan {et~al.}(2013)McQuillan, Mazeh, \&
  Aigrain}]{McQuillan2013a}
McQuillan, A., Mazeh, T., \& Aigrain, S. 2013, ApJ, 775, L11.
\newblock \url{http://adsabs.harvard.edu/abs/2013ApJ...775L..11M}

\bibitem[{McQuillan {et~al.}(2014)McQuillan, Mazeh, \& Aigrain}]{McQuillan2014}
---. 2014, ApJS, 211, 24.
\newblock \url{http://adsabs.harvard.edu/abs/2014ApJS..211...24M}

\bibitem[{Meibom {et~al.}(2015)Meibom, Barnes, Platais, Gilliland, Latham, \&
  Mathieu}]{Meibom2015}
Meibom, S., Barnes, S.~A., Platais, I., {et~al.} 2015, Nature, 517, 589.
\newblock \url{http://adsabs.harvard.edu/abs/2015Natur.517..589M}

\bibitem[{Meibom {et~al.}(2011{\natexlab{a}})Meibom, Mathieu, Stassun,
  Liebesny, \& Saar}]{Meibom2011a}
Meibom, S., Mathieu, R.~D., Stassun, K.~G., Liebesny, P., \& Saar, S.~H.
  2011{\natexlab{a}}, ApJ, 733, 115.
\newblock \url{http://adsabs.harvard.edu/abs/2011ApJ...733..115M}

\bibitem[{Meibom {et~al.}(2011{\natexlab{b}})Meibom, Barnes, Latham, Batalha,
  Borucki, Koch, Basri, Walkowicz, Janes, Jenkins, Van~Cleve, Haas, Bryson,
  Dupree, Furesz, Szentgyorgyi, Buchhave, Clarke, Twicken, \&
  Quintana}]{Meibom2011b}
Meibom, S., Barnes, S.~A., Latham, D.~W., {et~al.} 2011{\natexlab{b}}, ApJL,
  733, L9.
\newblock \url{http://adsabs.harvard.edu/abs/2011ApJ...733L...9M}

\bibitem[{Metcalfe \& Egeland(2019)}]{Metcalfe2019}
Metcalfe, T.~S., \& Egeland, R. 2019, ApJ, 871, 39.
\newblock \url{http://adsabs.harvard.edu/abs/2019ApJ...871...39M}

\bibitem[{Montet {et~al.}(2017)Montet, Tovar, \& Foreman-Mackey}]{Montet2017}
Montet, B.~T., Tovar, G., \& Foreman-Mackey, D. 2017, ApJ, 851, 116.
\newblock \url{http://adsabs.harvard.edu/abs/2017ApJ...851..116M}

\bibitem[{Mosser {et~al.}(2009)Mosser, Baudin, Lanza, Hulot, Catala, Baglin, \&
  Auvergne}]{Mosser2009}
Mosser, B., Baudin, F., Lanza, A.~F., {et~al.} 2009, A\&A, 506, 245.
\newblock \url{http://adsabs.harvard.edu/abs/2009A%26A...506..245M}

\bibitem[{Mosser {et~al.}(2018)Mosser, Gehan, Belkacem, Samadi, Michel, \&
  Goupil}]{Mosser2018}
Mosser, B., Gehan, C., Belkacem, K., {et~al.} 2018, A\&A, 618, A109.
\newblock \url{http://adsabs.harvard.edu/abs/2018A%26A...618A.109M}

\bibitem[{Murphy {et~al.}(2019)Murphy, Hey, Van~Reeth, \& Bedding}]{Murphy2019}
Murphy, S.~J., Hey, D., Van~Reeth, T., \& Bedding, T.~R. 2019, MNRAS, 485,
  2380.
\newblock \url{https://academic.oup.com/mnras/article/485/2/2380/5368359}

\bibitem[{Nemec {et~al.}(2013)Nemec, Cohen, Ripepi, Derekas, Moskalik, Sesar,
  Chadid, \& Bruntt}]{Nemec2013}
Nemec, J.~M., Cohen, J.~G., Ripepi, V., {et~al.} 2013, ApJ, 773, 181.
\newblock \url{http://adsabs.harvard.edu/abs/2013ApJ...773..181N}

\bibitem[{Nemec {et~al.}(2011)Nemec, Smolec, Benk{\H o}, Moskalik, Kolenberg,
  Szab{\'o}, Kurtz, Bryson, Guggenberger, Chadid, Jeon, Kunder, Layden,
  Kinemuchi, Kiss, Poretti, Christensen-Dalsgaard, Kjeldsen, Caldwell, Ripepi,
  Derekas, Nuspl, Mullally, Thompson, \& Borucki}]{Nemec2011}
Nemec, J.~M., Smolec, R., Benk{\H o}, J.~M., {et~al.} 2011, MNRAS, 417, 1022.
\newblock \url{http://adsabs.harvard.edu/abs/2011MNRAS.417.1022N}

\bibitem[{Nielsen {et~al.}(2013)Nielsen, Gizon, Schunker, \&
  Karoff}]{Nielsen2013}
Nielsen, M.~B., Gizon, L., Schunker, H., \& Karoff, C. 2013, A\&A, 557, L10.
\newblock \url{http://adsabs.harvard.edu/abs/2013A%26A...557L..10N}

\bibitem[{Paxton {et~al.}(2018)Paxton, Schwab, Bauer, Bildsten, Blinnikov,
  Duffell, Farmer, Goldberg, Marchant, Sorokina, Thoul, Townsend, \&
  Timmes}]{Paxton2018}
Paxton, B., Schwab, J., Bauer, E.~B., {et~al.} 2018, ApJS, 234, 34.
\newblock \url{https://ui.adsabs.harvard.edu/abs/2018ApJS..234...34P/abstract}

\bibitem[{Pedregosa {et~al.}(2011)Pedregosa, Varoquaux, Gramfort, Michel,
  Thirion, Grisel, Blondel, Prettenhofer, Weiss, Dubourg, Vanderplas, Passos,
  Cournapeau, Brucher, Perrot, \& Duchesnay}]{scikit-learn}
Pedregosa, F., Varoquaux, G., Gramfort, A., {et~al.} 2011, J. Mach. Learn.
  Res., 12, 2825.
\newblock \url{https://scikit-learn.org}

\bibitem[{Pires {et~al.}(2015)Pires, Mathur, Garc{\'i}a, Ballot, Stello, \&
  Sato}]{Pires2015}
Pires, S., Mathur, S., Garc{\'i}a, R.~A., {et~al.} 2015, A\&A, 574, A18.
\newblock \url{http://adsabs.harvard.edu/abs/2015A%26A...574A..18P}

\bibitem[{Reinhold {et~al.}(2019)Reinhold, Bell, Kuszlewicz, Hekker, \&
  Shapiro}]{Reinhold2019}
Reinhold, T., Bell, K.~J., Kuszlewicz, J., Hekker, S., \& Shapiro, A.~I. 2019,
  A\&A, 621, A21.
\newblock \url{http://adsabs.harvard.edu/abs/2019A%26A...621A..21R}

\bibitem[{Reinhold \& Hekker(2020)}]{Reinhold2020}
Reinhold, T., \& Hekker, S. 2020, A\&A, 635, A43.
\newblock \url{http://adsabs.harvard.edu/abs/2020A%26A...635A..43R}

\bibitem[{Ricker {et~al.}(2014)Ricker, Winn, Vanderspek, Latham, Bakos, Bean,
  Berta-Thompson, Brown, Buchhave, Butler, Butler, Chaplin, Charbonneau,
  Christensen-Dalsgaard, Clampin, Deming, Doty, De~Lee, Dressing, Dunham, Endl,
  Fressin, Ge, Henning, Holman, Howard, Ida, Jenkins, Jernigan, Johnson,
  Kaltenegger, Kawai, Kjeldsen, Laughlin, Levine, Lin, Lissauer, MacQueen,
  Marcy, McCullough, Morton, Narita, Paegert, Palle, Pepe, Pepper, Quirrenbach,
  Rinehart, Sasselov, Sato, Seager, Sozzetti, Stassun, Sullivan, Szentgyorgyi,
  Torres, Udry, \& Villasenor}]{Ricker2014}
Ricker, G.~R., Winn, J.~N., Vanderspek, R., {et~al.} 2014, Proceedings of the
  SPIE, 9143, 914320, conference Name: Space Telescopes and Instrumentation
  2014: Optical, Infrared, and Millimeter Wave.
\newblock \url{http://adsabs.harvard.edu/abs/2014SPIE.9143E..20R}

\bibitem[{Salabert {et~al.}(2017)Salabert, Garc{\'i}a, Jim{\'e}nez, Bertello,
  Corsaro, \& Pall{\'e}}]{Salabert2017}
Salabert, D., Garc{\'i}a, R.~A., Jim{\'e}nez, A., {et~al.} 2017, A\&A, 608,
  A87.
\newblock \url{http://adsabs.harvard.edu/abs/2017A%26A...608A..87S}

\bibitem[{Salabert {et~al.}(2016)Salabert, Garc{\'i}a, Beck, Egeland,
  Pall{\'e}, Mathur, Metcalfe, do~Nascimento, Ceillier, Andersen, \&
  Trivi{\~n}o~Hage}]{Salabert2016a}
Salabert, D., Garc{\'i}a, R.~A., Beck, P.~G., {et~al.} 2016, A\&A, 596, A31.
\newblock \url{http://adsabs.harvard.edu/abs/2016A%26A...596A..31S}

\bibitem[{Santos {et~al.}(2019)Santos, Garc{\'i}a, Mathur, Bugnet, Saders,
  Metcalfe, Simonian, \& Pinsonneault}]{Santos2019a}
Santos, A. R.~G., Garc{\'i}a, R.~A., Mathur, S., {et~al.} 2019, ApJS, 244, 21.
\newblock \url{https://doi.org/10.3847%2F1538-4365%2Fab3b56}

\bibitem[{Simonian {et~al.}(2019)Simonian, Pinsonneault, \&
  Terndrup}]{Simonian2019}
Simonian, G. V.~A., Pinsonneault, M.~H., \& Terndrup, D.~M. 2019, ApJ, 871,
  174.
\newblock \url{http://adsabs.harvard.edu/abs/2019ApJ...871..174S}

\bibitem[{Skumanich(1972)}]{Skumanich1972}
Skumanich, A. 1972, ApJ, 171, 565.
\newblock \url{http://adsabs.harvard.edu/abs/1972ApJ...171..565S}

\bibitem[{Smith {et~al.}(2012)Smith, Stumpe, Van~Cleve, Jenkins, Barclay,
  Fanelli, Girouard, Kolodziejczak, McCauliff, Morris, \& Twicken}]{Smith2012}
Smith, J.~C., Stumpe, M.~C., Van~Cleve, J.~E., {et~al.} 2012, PASP, 124, 1000.
\newblock \url{http://adsabs.harvard.edu/abs/2012PASP..124.1000S}

\bibitem[{Stumpe {et~al.}(2012)Stumpe, Smith, Van~Cleve, Twicken, Barclay,
  Fanelli, Girouard, Jenkins, Kolodziejczak, McCauliff, \& Morris}]{Stumpe2012}
Stumpe, M.~C., Smith, J.~C., Van~Cleve, J.~E., {et~al.} 2012, PASP, 124, 985.
\newblock \url{http://adsabs.harvard.edu/abs/2012PASP..124..985S}

\bibitem[{Torrence \& Compo(1998)}]{Torrence1998}
Torrence, C., \& Compo, G.~P. 1998, Bull. Amer. Meteor. Soc., 79, 61.
\newblock
  \url{https://journals.ametsoc.org/doi/abs/10.1175/1520-0477%281998%29079%3C0061%3AAPGTWA%3E2.0.CO%3B2}

\bibitem[{Uytterhoeven {et~al.}(2011)Uytterhoeven, Moya, Grigahc{\`e}ne, Guzik,
  Guti{\'e}rrez-Soto, Smalley, Handler, Balona, Niemczura, Fox~Machado,
  Benatti, Chapellier, Tkachenko, Szab{\'o}, Su{\'a}rez, Ripepi, Pascual,
  Mathias, Mart{\'i}n-Ru{\'i}z, Lehmann, Jackiewicz, Hekker, Gruberbauer,
  Garc{\'i}a, Dumusque, D{\'i}az-Fraile, Bradley, Antoci, Roth, Leroy, Murphy,
  De~Cat, Cuypers, Kjeldsen, Christensen-Dalsgaard, Breger, Pigulski, Kiss,
  Still, Thompson, \& van Cleve}]{Uytterhoeven2011}
Uytterhoeven, K., Moya, A., Grigahc{\`e}ne, A., {et~al.} 2011, A\&A, 534, A125.
\newblock
  \url{https://ui.adsabs.harvard.edu/abs/2011A%26A...534A.125U/abstract}

\bibitem[{Van~Reeth {et~al.}(2018)Van~Reeth, Mombarg, Mathis, Tkachenko,
  Fuller, Bowman, Buysschaert, Johnston, Garc{\'i}a~Hern{\'a}ndez, Goldstein,
  Townsend, \& Aerts}]{vanReeth2018}
Van~Reeth, T., Mombarg, J. S.~G., Mathis, S., {et~al.} 2018, A\&A, 618, A24.
\newblock \url{http://adsabs.harvard.edu/abs/2018A%26A...618A..24V}

\bibitem[{van Saders {et~al.}(2016)van Saders, Ceillier, Metcalfe,
  Silva~Aguirre, Pinsonneault, Garc{\'i}a, Mathur, \& Davies}]{vanSaders2016}
van Saders, J.~L., Ceillier, T., Metcalfe, T.~S., {et~al.} 2016, Nature, 529,
  181.
\newblock \url{http://adsabs.harvard.edu/abs/2016Natur.529..181V}

\bibitem[{van Saders {et~al.}(2019)van Saders, Pinsonneault, \&
  Barbieri}]{vanSaders2019}
van Saders, J.~L., Pinsonneault, M.~H., \& Barbieri, M. 2019, ApJ, 872, 128.
\newblock \url{http://adsabs.harvard.edu/abs/2019ApJ...872..128V}

\bibitem[{Vaughan {et~al.}(1981)Vaughan, Baliunas, Middelkoop, Hartmann,
  Mihalas, Noyes, \& Preston}]{Vaughan1981}
Vaughan, A.~H., Baliunas, S.~L., Middelkoop, F., {et~al.} 1981, ApJ, 250, 276.
\newblock \url{http://adsabs.harvard.edu/abs/1981ApJ...250..276V}

\bibitem[{Virtanen {et~al.}(2020)Virtanen, Gommers, Oliphant, Haberland, Reddy,
  Cournapeau, Burovski, Peterson, Weckesser, Bright, {van der Walt}, Brett,
  Wilson, Millman, Mayorov, Nelson, Jones, Kern, Larson, Carey, Polat, Feng,
  Moore, {VanderPlas}, Laxalde, Perktold, Cimrman, Henriksen, Quintero, Harris,
  Archibald, Ribeiro, Pedregosa, {van Mulbregt}, \& {SciPy 1.0
  Contributors}}]{2020SciPy-NMeth}
Virtanen, P., Gommers, R., Oliphant, T.~E., {et~al.} 2020, Nature Methods, 17,
  261

\bibitem[{{W}es {M}c{K}inney(2010)}]{mckinney-proc-scipy-2010}
{W}es {M}c{K}inney. 2010, in {P}roceedings of the 9th {P}ython in {S}cience
  {C}onference, ed. {S}t\'efan van~der {W}alt \& {J}arrod {M}illman, 56 -- 61.
\newblock
  \url{https://conference.scipy.org/proceedings/scipy2010/mckinney.html}

\end{thebibliography}

\appendix

\section{ROOSTER's performance for hot solar-like stars}\label{app:MLtrainingset}

\ml\ (Sect.~\ref{sec:ML}) was developed and validated in \citet{Breton2021} with the target sample of Paper~I. \citet{Breton2021} performed a training loop with 100 realizations. In each realization, 75\% of the targets are randomly selected for the training set, while the remainder 25\% constitutes the test set. By performing a training loop rather than a single training, one can compute the mean classification ratio for each star. As follows, \ml\ is able to classify all the targets in the training set.

In this section, we discuss \ml's results for the full training sample of the current analysis (see Sect.~\ref{sec:training}), which, in addition to the cooler stars of Paper~I, includes hotter targets as well. From \citet{Breton2021} to the current analysis, we have more than doubled the size of the training set. This way \ml\ is trained with targets of spectral types from mid-F to M (Fig.~\ref{fig:hr_training}). This increment is motivated by the different behavior observed in F stars in comparison to cooler solar-like stars in terms of rotational signature (see discussion in Sect.~\ref{sec:conclusions}). 

\begin{figure}[h]
    \centering
    \includegraphics[width=0.65\hsize]{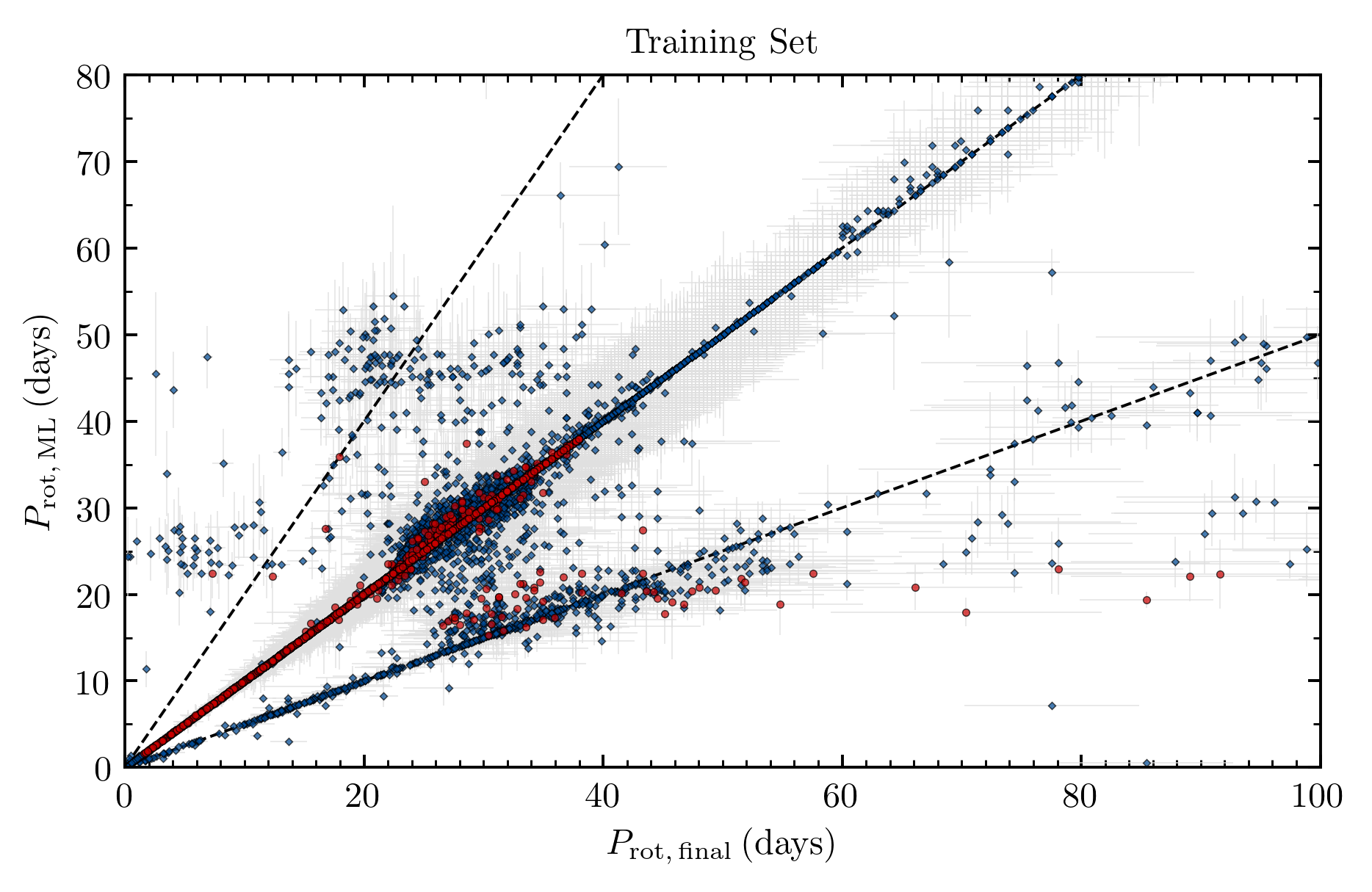}\vspace{-0.2cm}
    \caption{Same as in Fig.~\ref{fig:final_ML} but for the training set (Sect.~\ref{sec:training}). The blue diamonds mark the targets that would be selected for visual inspection or for an automatic change in the filter choice following the criteria described in Sect.~\ref{sec:vischecks}.}
    \label{fig:MLtrainingset}
\end{figure}

Figure~\ref{fig:MLtrainingset} compares the rotation periods selected by \ml\ (\protml) and the correct \prot. The blue diamonds highlight the targets that would be selected for visual inspection or for an automatic change in the filter choice according to the selection criteria described in Sect.~\ref{sec:vischecks}. The \prot\ values agree within 15\% for $\sim 95.3\%$ of the targets. From the targets in disagreement, $\sim94.6\%$ would be selected for visual inspection (blue diamonds) and, therefore, corrected. The global accuracy of \ml\ for the training set comprised of mid-F to M stars is 95.3\%. Finally, \ml\ only selects \prot\ for two targets that are found not to have rotational modulation and only misses \prot\ for the targets with missing parameters, which would be selected for visual inspection following the procedure in Sect.~\ref{sec:vischecks}. 

\section{Main-sequence F stars and Kraft break}\label{app:Fstars}

\begin{figure}[h]
    \centering
    \includegraphics[width=0.495\hsize]{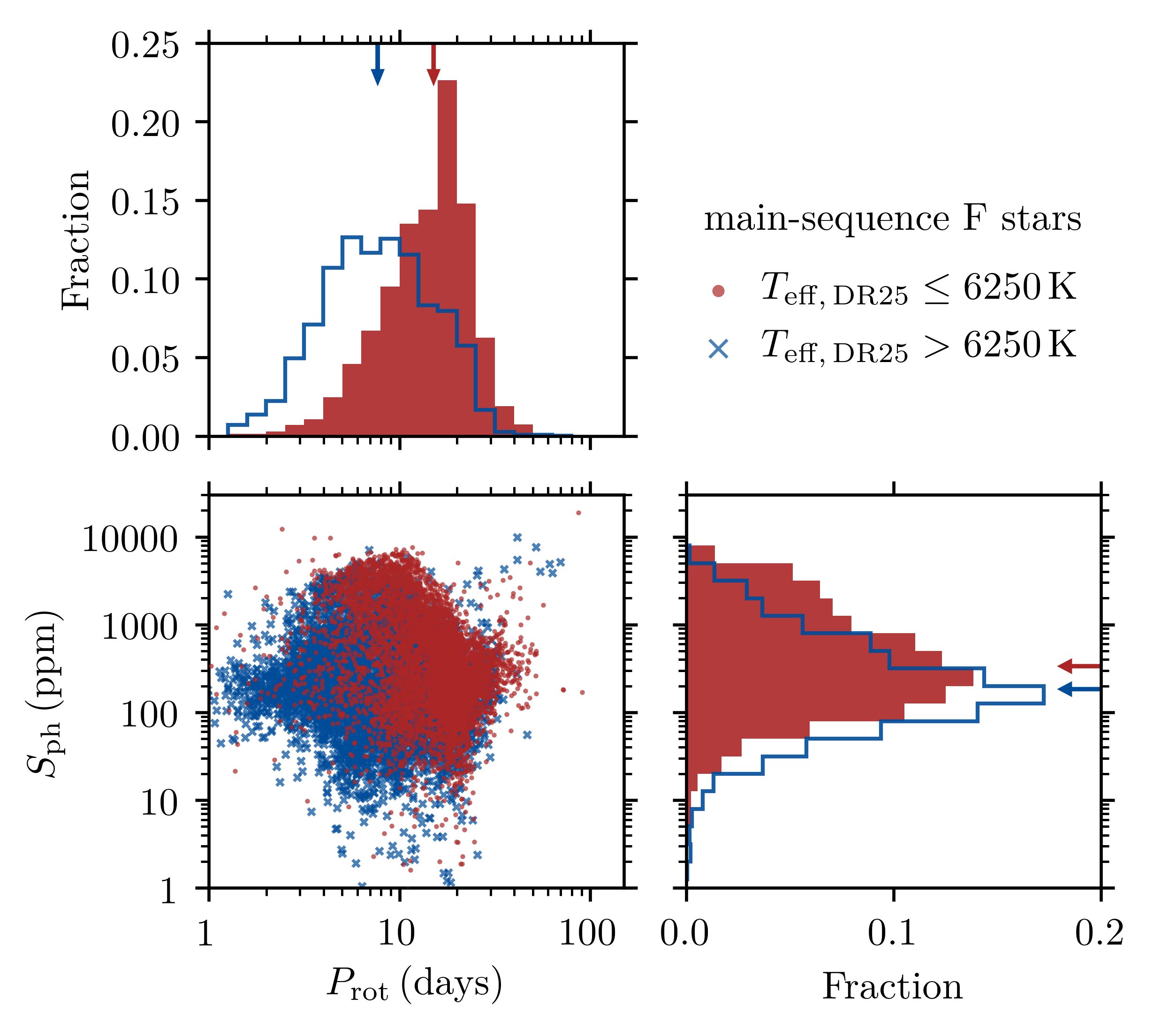}
    \includegraphics[width=0.495\hsize]{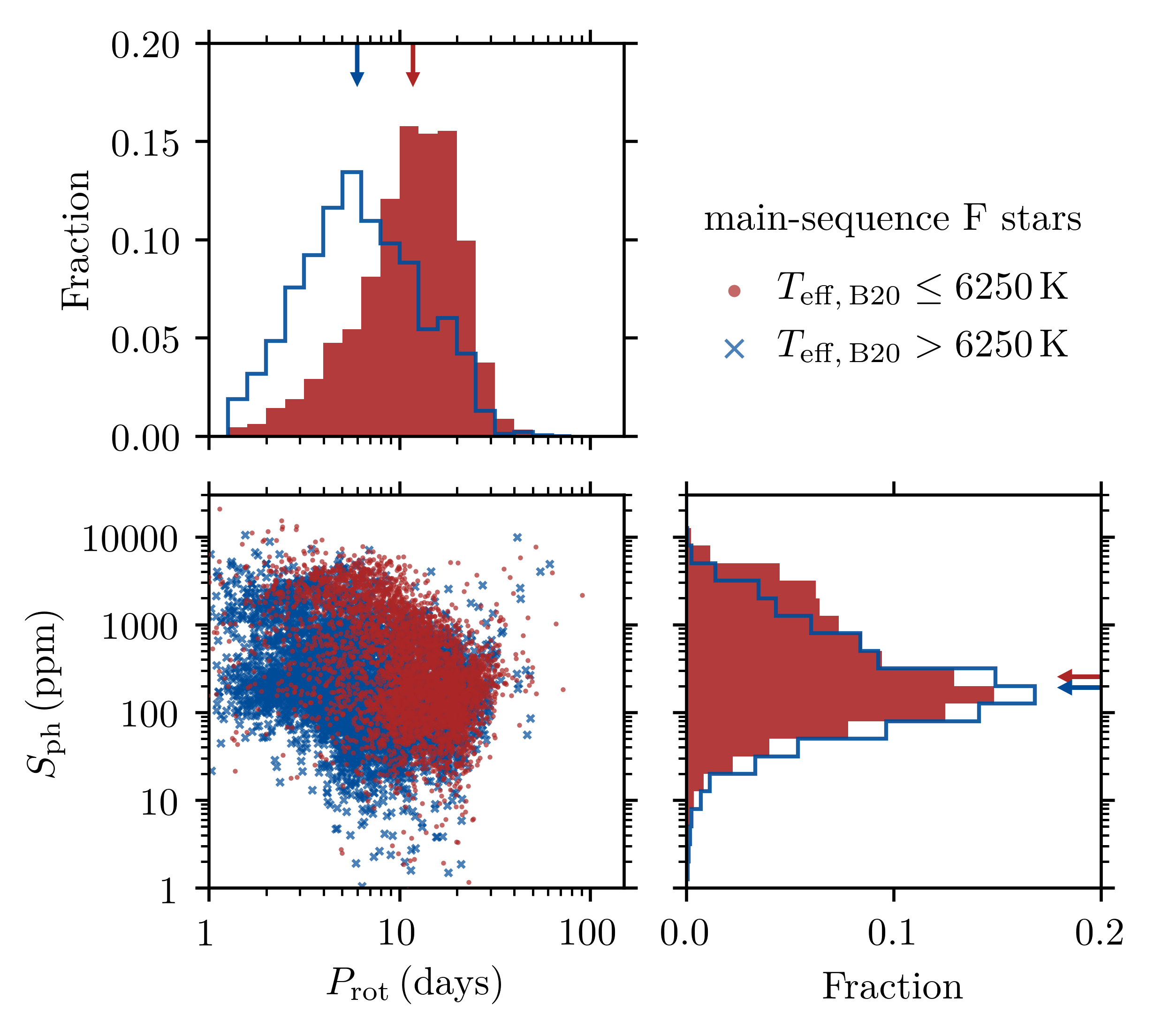}\vspace{-0.2cm}
    \caption{\sph\ and \prot\ distributions for the main-sequence F~stars in subsample~I below (red) and above (blue) the Kraft break ($T_\text{eff}=6250$ K). The left panels consider the stellar properties from DR25, while the right-hand side panels consider the B20 catalog. The arrows mark the median values of the distribution.}
    \label{fig:sphprot_F}
\end{figure}

The rotation period is observed to decrease generally with increasing effective temperature (Fig.~\ref{fig:protteff}). F stars are then typically fast rotators. For main-sequence GKM stars (Fig.~\ref{fig:protsph}; and Fig.~9 in Paper~I), faster rotators are found to be photometrically more active than slower rotators. However, for main-sequence F stars and subgiants there is a group of weakly active fast rotators. For subgiants, it is clear that those correspond to the hottest subgiant stars considered in this work (Figs.~\ref{fig:protteff} and \ref{fig:protsph}). Figure~\ref{fig:sphprot_F} shows the \sph-\prot\ diagram for the main-sequence F stars expected to be below (red) and above (blue) the Kraft break \citep{kraft_studies_1967}. Note that the red data points are overplotted. The left panels show the results based on the stellar properties from DR25, while the right-hand panels show the results based on B20 \citep{Berger2020}, where F stars usually have lower \teff\ compared to DR25. Most of the fast rotating F stars with small \sph\ values are stars expected to be above the Kraft break. Additionally to the \teff\ difference between the two catalogs, there is an associated uncertainty and the effect from metallicity on the convection properties, which may also contribute to the scatter in these diagrams.

\section{Slow-rotating subgiant stars}\label{app:subG}

\begin{figure}[h]
    \centering
    \includegraphics[width=\hsize]{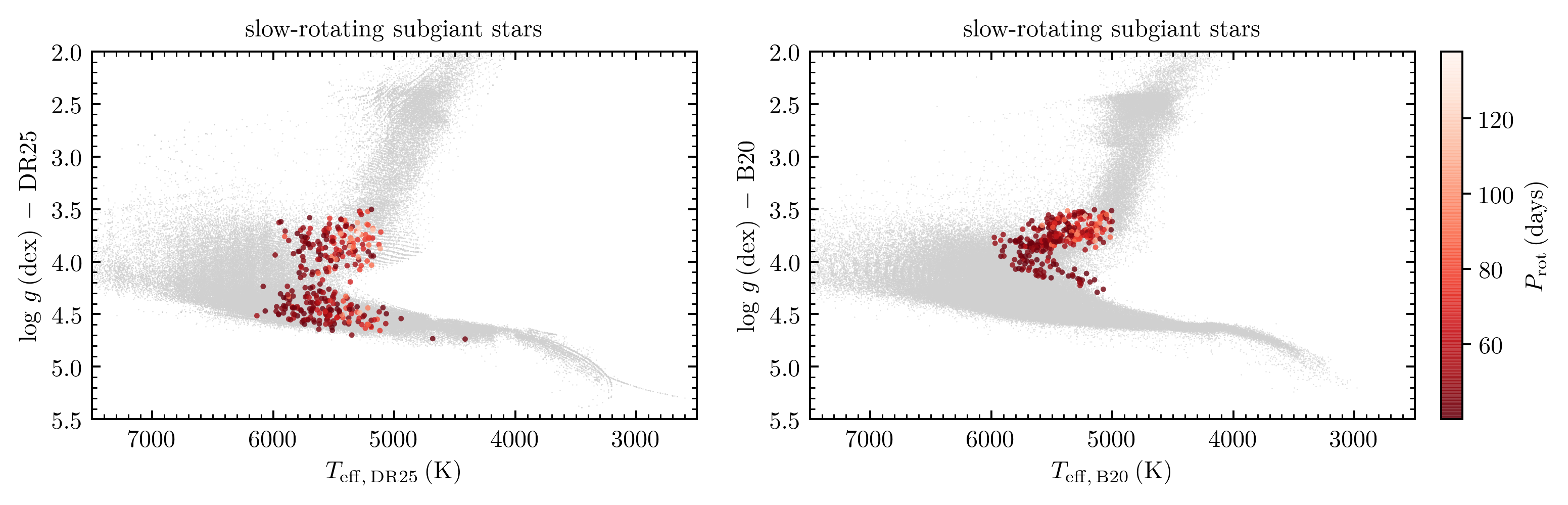}\vspace{-0.2cm}
    \caption{Slow-rotating subgiant stars (selected according to B20; shades of red) in the \logg-\teff\ diagram according to DR25 (left) and B20 (right). The data points are color-coded by \prot. For reference the gray dots depict all the targets in the DR25 and B20.}
    \label{fig:hr_subG_slow}
\end{figure}

While the \prot\ values for most of the subgiant stars are consistent with the \prot\ distribution for the main-sequence stars of similar \teff, there is a group of slow-rotating stars (Fig.~\ref{fig:protteff}), which are located above the upper edge of the \prot\ distribution. We select the targets with $5000\leq T_\text{eff,\,B20}\leq6000$ K and $P_\text{rot}>40$ days. Figure~\ref{fig:hr_subG_slow} shows where these slow-rotating subgiants stars are located in the \logg-\teff\ diagram according to the stellar properties from DR25 (left) and B20 (right). The slowest targets (lighter colors) tend to be more evolved targets, in both DR25 and B20, relative to the target sample. However, part of slow-rotating subgiants are in the main sequence according to DR25. 

\pagebreak
\section{Full target sample}\label{app:allB20}

\begin{figure}[h]
    \centering
    \includegraphics[width=\hsize]{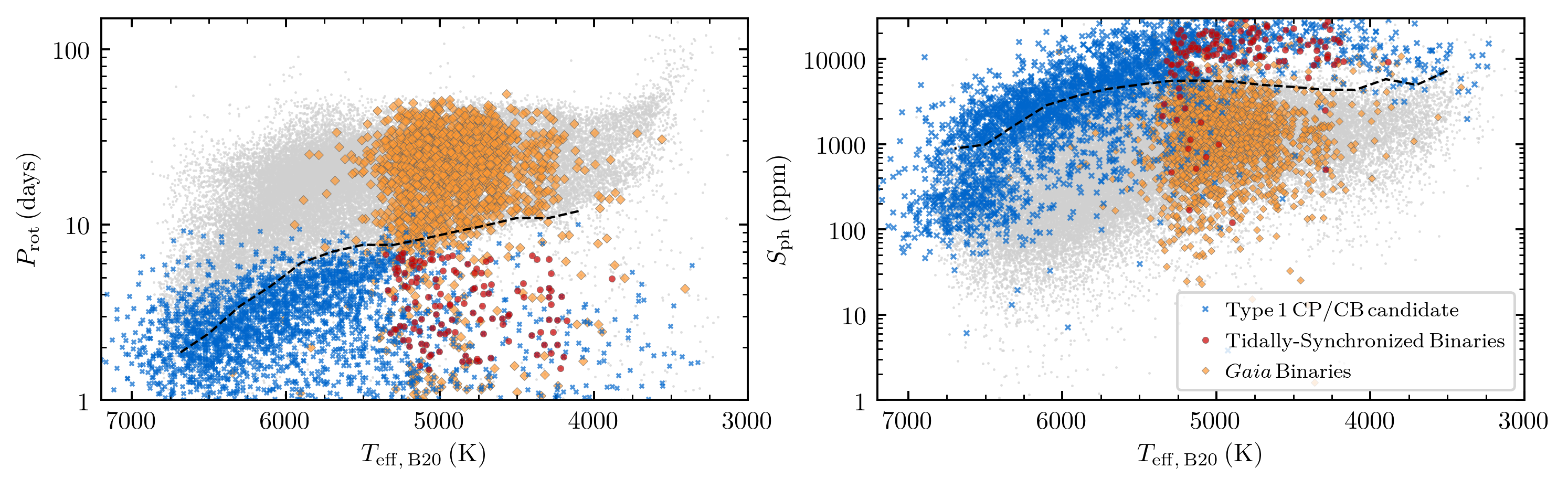}
    \caption{\prot\ (left) and \sph\ (right) as a function of \teff\ for: the \cpcbi\ candidates (blue crosses); synchronized binaries \citep[red circles;][]{Simonian2019}; and {\it Gaia} binaries \citep[yellow diamonds;][]{Berger2018}. For reference the gray data points show the main-sequence stars in Figs.~\ref{fig:protteff} and \ref{fig:sphteff} and the dashed black lines indicate the lower edge of the \prot\ distribution (left) and upper edge of the \sph\ distribution (right).}
    \label{fig:protsphteff_cpcb}
\end{figure}

Figure~\ref{fig:protsphteff_cpcb} highlights the \cpcbi\ candidates, tidally-synchronized binaries \citep{Simonian2019}, and {\it Gaia} binaries \citep{Berger2018} in the \prot-\teff\ and \sph-\teff\ diagrams for the targets with \prot\ estimate in Paper~I and in this work. The overlap between the target sample of the current paper and the binaries identified by \citet{Simonian2019} and \citet{Berger2018} is very small, with only 8 and 125 targets, respectively. The tidally-synchronized binaries tend to have larger \sph\ values and short periods in comparison to the targets with similar \teff. The {\it Gaia} binaries have a \prot\ and \sph\ distribution more consistent with those of the, presumably, single targets. The targets flagged as \cpcbi\ candidates have generally short periods and large \sph. In particular, they are located at or beyond the lower edge of the \prot\ distribution and the upper edge of the \sph\ distribution. To guide the eye, the dashed lines in Fig.~\ref{fig:protsphteff_cpcb} show the lower (5\% percentile) and upper (95\% percentile) edges of the \prot\ and \sph\ distribution, respectively. The cooler end of the \prot\ edge was removed because of its erratic behaviour due to small sample size. Nevertheless, the gray data points shows the results for the main-sequence stars in Figs.~\ref{fig:protteff} and \ref{fig:sphteff}.

As concluded in Paper~I, the tidally-synchronized binaries and the \cpcbi\ candidates tend to occupy the same parameter space, being typically characterized by very large \sph\ and short \prot. Note however that data, methodology, and properties studied in \citet{Simonian2019} are distinct from those of this work and Paper~I.
The targets flagged as \cpcbi\ candidates were first identified in Paper~I during the visual inspection. The behaviour of the brightness variations and the respective rotation diagnostics appeared to be distinct from the remainder of the solar-like stars with rotational modulation. These targets exhibit large-amplitude brightness variations, leading to large \sph\ values. The ``rotational" modulation shows fast and stable beating patterns throughout the time-series, which also leads to, for example, beating in the ACF. Finally, these targets often show a large number of visible harmonics of the rotation period. Part of these targets were found to be tidally-synchronized binaries by \citet{Simonian2019}, who concluded that the rapid-rotating regime in \kep\ observations is dominated by binary systems. The overlap and the similarities between the \cpcbi\ candidates and the tidally-synchronized binaries suggested that \cpcbi\ candidates might be indeed binaries, while the signal may still be related to rotational modulation. For the current work, \ml\ was trained to flag these targets. As discussed in \citet{Breton2021}, \ml\ tends to flag more targets than those flagged by visual inspection. This may suggest that \ml\ is flagging targets that are not \cpcbi\ candidates. Nevertheless, we advise caution when dealing with these targets.

\begin{figure}[h]
    \centering
    \includegraphics[width=\hsize]{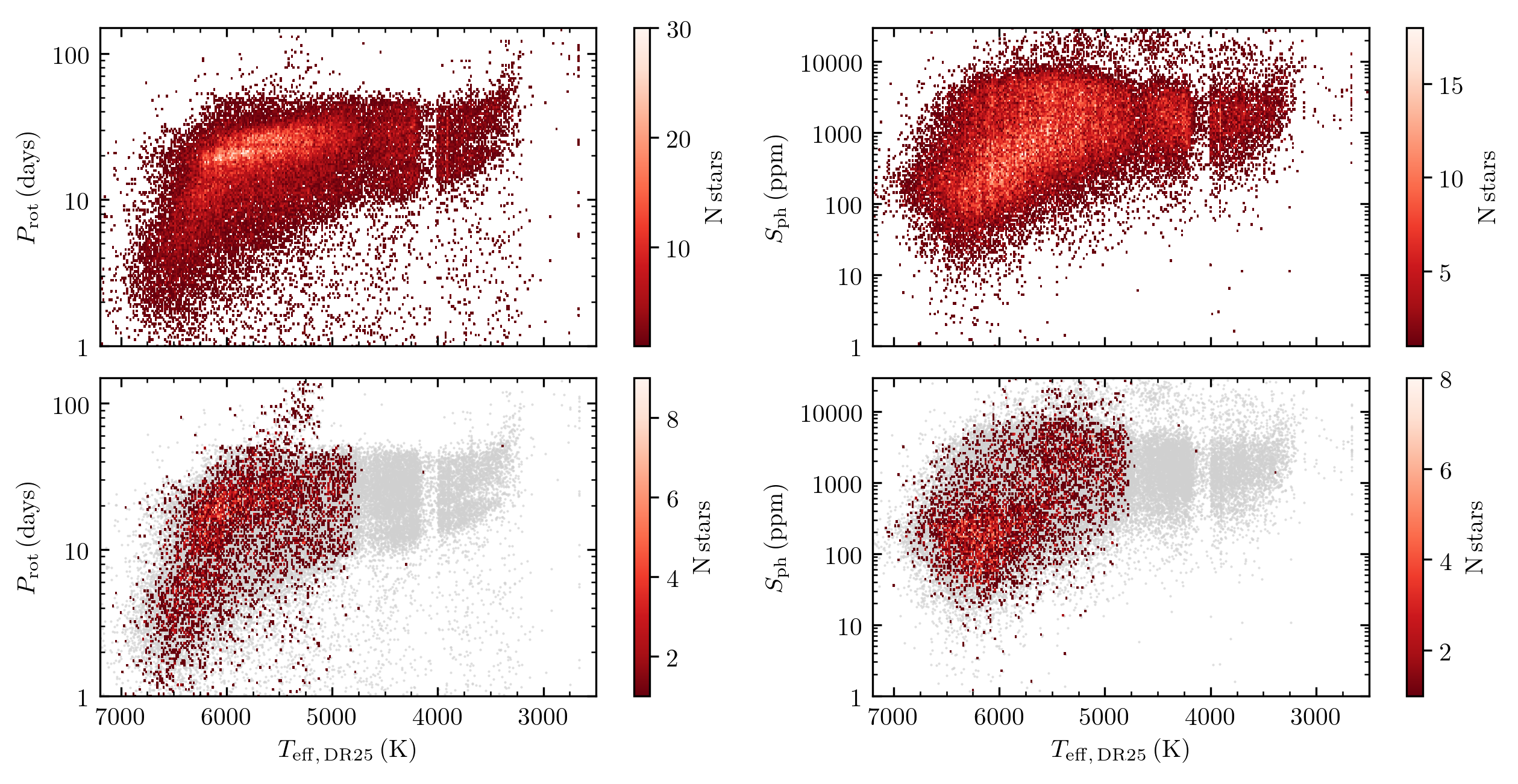}
    \caption{Same as in Figs.~\ref{fig:protteff} and \ref{fig:sphteff} but considering the stellar properties from DR25 (\logg\ is used to split subgiant and main-sequence stars).}
    \label{fig:protsphteffDR25}
\end{figure}

Figure \ref{fig:protsphteffDR25} shows the same as Figs.~\ref{fig:sphteff} and \ref{fig:protsph}  but for the stellar properties of DR25, where subgiant and main-sequence stars are separated according to \logg\ from DR25 (in this figure). As discussed in B20, the effective temperatures of M stars (to be improved in a forthcoming work; see B20) are overestimated in comparison with DR25. Thus, in Fig.~\ref{fig:protsphteffDR25} M stars are located at cooler \teff. The slow-rotating subgiants have hotter \teff\ in DR25 than in B20, while F stars are also shifted towards hotter temperatures. Note that the \teff\ gaps in DR25 are due to artifacts in the stellar properties catalog.

\begin{figure}[h]
    \centering
    \includegraphics[width=\hsize]{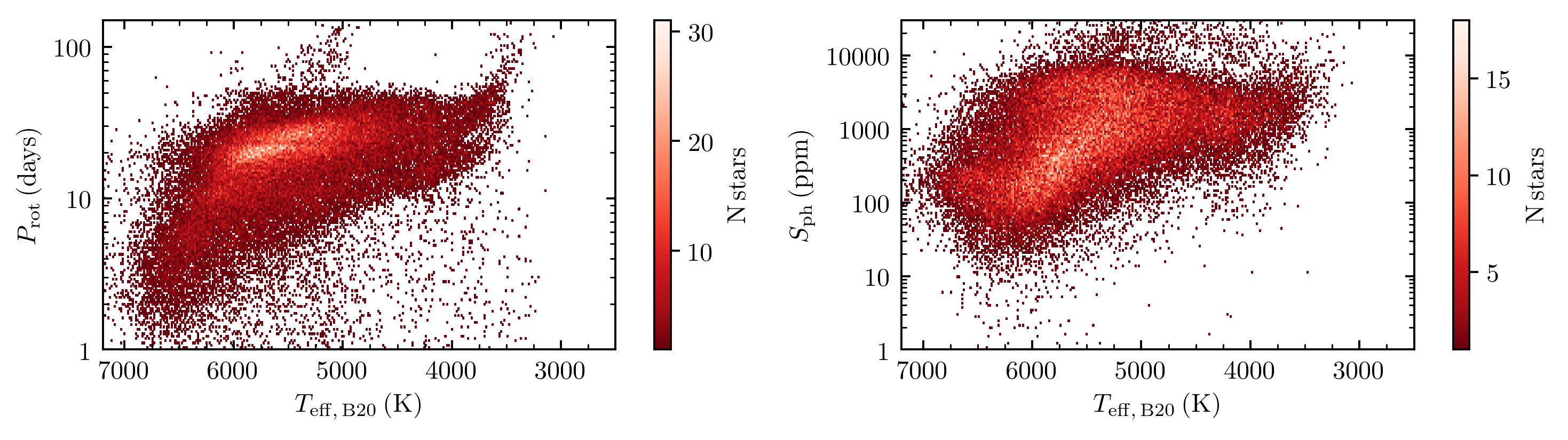}
    \caption{\prot\ (left) and \sph\ (right) as a function of \teff\ for all targets with \prot\ estimate from Paper I and subsamples I-III for comparison with Figs.~\ref{fig:protteff} and \ref{fig:sphteff}.}
    \label{fig:protsphteff_all}
\end{figure}

\begin{figure}[h]
    \centering
    \includegraphics[width=\hsize]{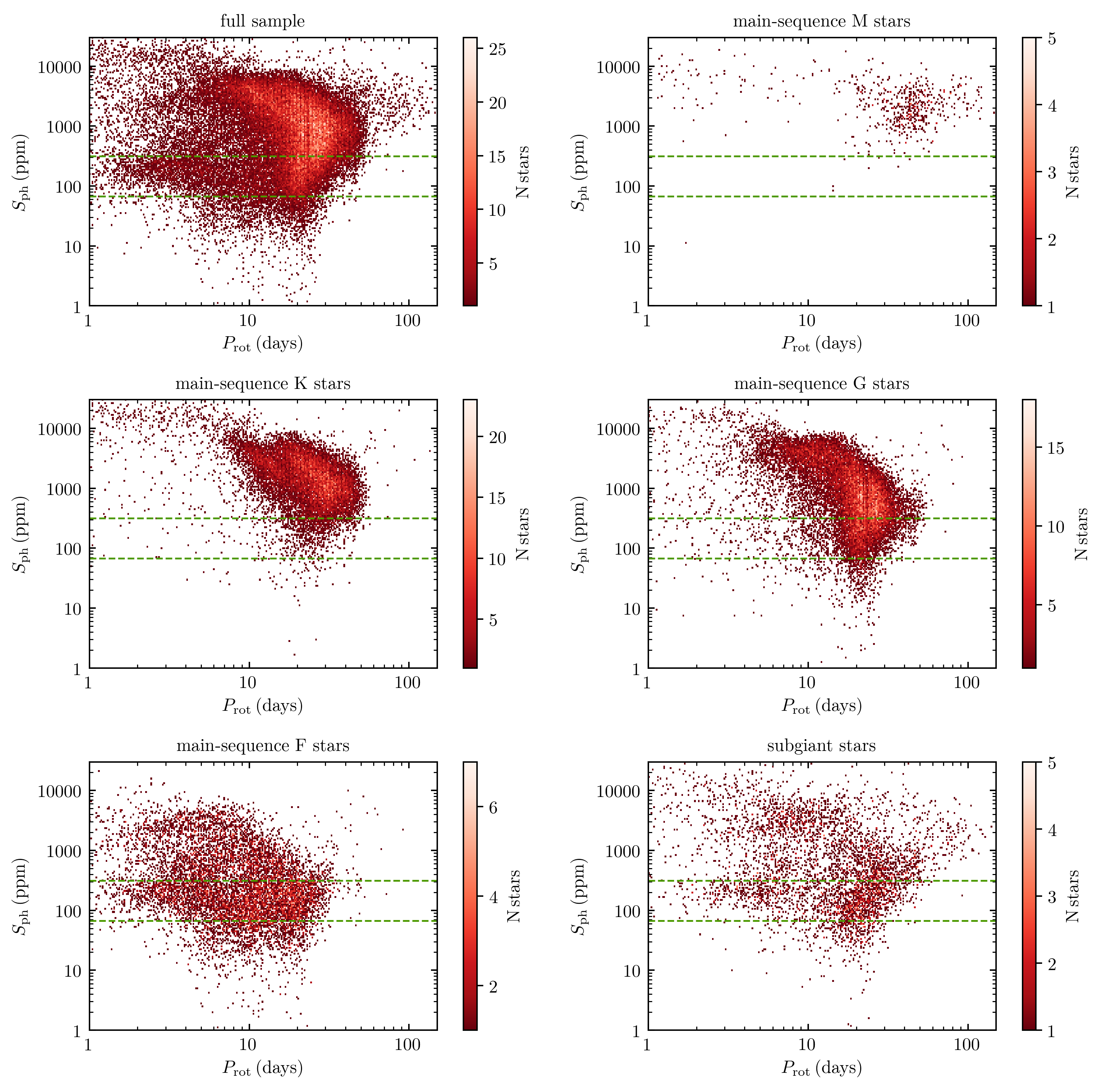}
    \caption{Same as Fig.\ref{fig:protsph} but for all targets with \prot\ estimate from Paper I and subsamples I-III (same targets as in Fig.~\ref{fig:protsphteff_all}). This includes \cpcbi\ candidates seen mainly in the left top corners of the panels (see Fig.~\ref{fig:protsphteff_cpcb}).}
    \label{fig:protsph_all}
\end{figure}

Figures \ref{fig:protsphteff_all} and \ref{fig:protsph_all} show the \prot-\teff\, \sph-\teff, and \sph-\prot\ diagrams for the full target sample of Paper~I and this work, including \cpcbi\ candidates and subsamples II and III, which were neglected in Figs.~\ref{fig:protteff} and \ref{fig:sphteff}. As discussed above, there is an increase of fast rotators, particularly with high \teff\, and an increase of large \sph\ values. In Figs.~\ref{fig:protsphteff_all} and \ref{fig:protsph_all}, the split of the targets in terms of spectral type and evolutionary state is made according to B20 stellar properties. Note that in Paper~I, we used DR25.

\begin{figure}
    \centering
    \includegraphics[width=0.45\hsize]{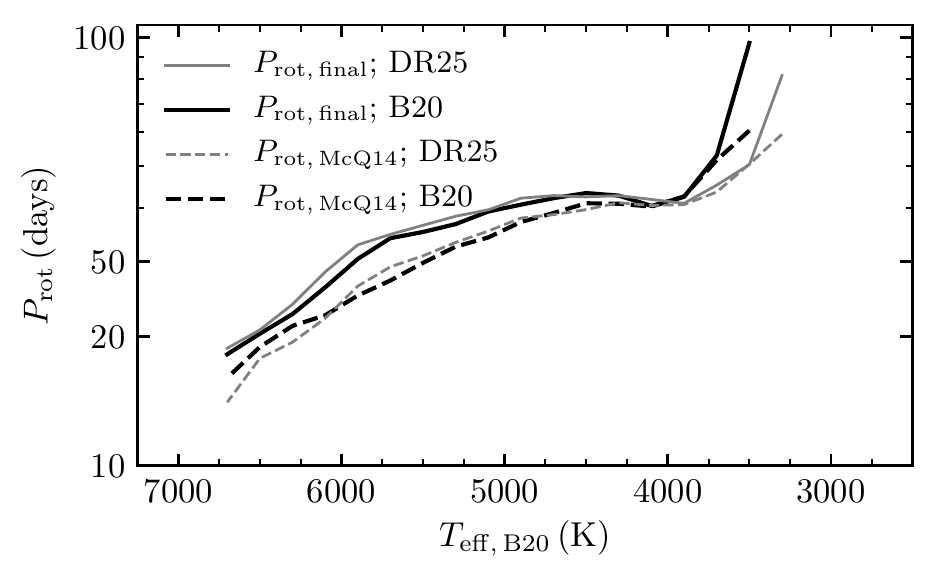}
    \caption{Same as in Fig.~\ref{fig:envelope} (thick black lines), but also for the \teff\ values from DR25 (thin gray lines). The solid lines concern the \prot\ estimates determined here and in Paper~I, while the dashed lines represent the results from McQ14.}
    \label{fig:envelopeB20}
\end{figure}

Finally, Fig.~\ref{fig:envelopeB20} compares the upper edge of the \prot\ distribution according to the \teff\ values in DR25 and B20 (for comparison with the simplified Fig.~\ref{fig:envelope}). The upper edge does not change significantly due to the different \teff\ estimates, in particular for stars hotter than 4000 K. For M stars, as mentioned above $T_\text{eff,\,B20}$ is systematically larger than $T_\text{eff,\,DR25}$. Independently on the stellar properties catalog, our \prot\ distribution is characterized by a larger number of slow rotators than that of McQ14.

\end{document}